\documentclass[aps,pre,preprint]{revtex4-2}
\usepackage{amsmath}
\usepackage{graphicx}
\usepackage{bbm}

\begin{document}
\title{Energy Band Structure of Multistream Quantum Electron System}
\author{M. Akbari-Moghanjoughi}
\affiliation{Faculty of Sciences, Department of Physics, Azarbaijan Shahid Madani University, 51745-406 Tabriz, Iran}

\begin{abstract}
In this paper, using the quantum multistream model, we develop a method to study the electronic band structure of plasmonic excitations in streaming electron gas with arbitrary degree of degeneracy. The multifluid quantum hydrodynamic model is used to obtain $N$-coupled pseudoforce differential equation system from which the energy band structure of plasmonic excitations is calculated. It is shown that inevitable appearance of energy bands separated by gaps can be due to discrete velocity filaments and their electrostatic mode coupling in the electron gas. Current model also provides an alternative description of collisionless damping and phase mixing, i.e., collective scattering phenomenon within the energy band gaps due to mode coupling between wave-like and particle-like oscillations. The quantum multistream model is further generalized to include virtual streams which is used to calculate the electronic band structure of one-dimensional plasmonic crystals. It is remarked that, unlike the empty lattice approximation in free electron model, energy band gaps exist in plasmon excitations due to the collective electrostatic interactions between electrons. It is also shown that the plasmonic band gap size at first Brillouin zone boundary maximizes at the reciprocal lattice vector, $G$, close to metallic densities. Furthermore, the electron-lattice binding and electron-phonon coupling strength effects on the electronic band structure are discussed. It is remarked that inevitable formation of energy band structure is a general characteristics of various electromagnetically and gravitationally coupled quantum multistream systems.
\end{abstract}

\date{\today}
\email[Corresponding author: ]{massoud2002@yahoo.com}
\maketitle
\newpage

\section{Introduction}

Plasmons are high frequency elementary quantized excitations of electron plasma oscillations \cite{chen,krall}. They play inevitable role in many fundamental properties of plasmas semiconductors and metallic nanoparticles from electric and heat transport phenomena to optical and dielectric response, etc. \cite{kit,ash}. Dynamics of these quantized electromagnetic quasiparticles make an ideal platform for miniaturization of ultrafast terahertz device communications \cite{umm}, where conventional integrated circuits fail to operate. They also have numerous other interesting applications in nanotechnology \cite{mark}, plasmonics \cite{gardner,man1,maier}, optoelectronics \cite{haug}, etc. for engineering low-dimensional nano-fabricated semiconductor industry \cite{hu,seeg,at}. Energy conversion by plasmons is an new way of solar power extraction due to its high efficiency in photovoltaic and catalytic devices. Use of the collective oscillations of electrons instead of single particles makes huge amount of energy extraction in an operation step in plasmonic solar devices \cite{cesar,jac}.In local surface plasmon resonance (LSPR) \cite{at2} process, the surface electrons, the so called hot electrons, are collectively excited by electromagnetic radiations in UV-VIS range generating a huge amount of energy transfer. The hot electron current are collected in an appropriate contacts of nanoparticle surfaces by an efficient electron collecting material like $TiO_2$ in Schottky configuration \cite{tian}.

Collective charge screening effect which manifests itself as the characteristic optical edge in metallic surfaces already have may applications in metallic alloys making then optically unique among other solids. Collective electron excitations rule almost every aspect of solid from optical to dielectric response in plasmas \cite{ichimaru1,ichimaru2,ichimaru3} and condensed matter. Recent infrared spectroscopic techniques shows that Low dimensional semiconductors \cite{yofee} such as gapped graphene also demonstrate interesting surface plasmon effects. The collective electron transport property of graphene makes it an ideal element for multilayer composite devices such as compact ultrafast switches, optical modulators, optical lattices, photodetectors, tandem solar cells and biosensors \cite{jian,fey,hugen}. The first theoretical development of the idea of collective electron excitations by Bohm and Pines dates back to mid-nineties, when they coined plasmon name for such excitations due to the long-range electromagnetic nature of interactions \cite{bohm,bohm1,bohm2,pines,levine}. The theoretical as well as experimental aspects of collective electron dynamics in quantum level has been the subject of intense investigations over the past few decades \cite{march,swada,kohn,fetter,mahan,pin}, due to its fundamental importance in many field of physics and chemistry.

Pioneering developments of quantum statistical and kinetic theories \cite{fermi,chandra,hoyle,klimontovich} had a long tradition furnishing a pavement for modern theories of quantum plasmas \cite{haas1,man0,haasbook,bonitz0}. Many interesting new aspects of collective quantum effects in astrophysical and laboratory plasmas has been recently investigated using quantum plasma theories \cite{se,sten,ses,brod1,mark1,man4,fhaas,scripta,stenf1,sm,michta,ali1,ali2,akbground}. The quantum kinetic theories like time-dependent density functional theories (TDFT) are, however, less analytic as compared to the quantum hydrodynamic analogues, due mostly to mathematical complexity which require large scale computational programming. Recent investigation reveals \cite{manew} that quantum hydrodynamic approaches based on the density functional formalism \cite{bonitz0} can reach beyond the previously thought kinetic limitations, such as the collisionless damping if accurately formulated. One of the most effective hydrodynamic formalism for studying the quantum aspects of plasmas is the Schr\"{o}dinger-Poisson model \cite{manfredi,hurst}, based on the Madelung quantum fluid theory which originally attempted for the single-electron quantum fluid modeling \cite{madelung}. It has been recently shown that the analytic investigation of linearized Schr\"{o}dinger-Poisson system for arbitrary degenerate electron gas provides routes to some novel quantum feature of collective plasmon excitations \cite{akbhd,akbquant}. In current study we use the multistream model in order to investigate the band structure plasmon excitations in streaming plasmas and plasmonic lattices.

\section{Mathematical Model}

Starting with a one dimensional collision-less multi-fluid quantum hydrodynamic model for electron gas with an arbitrary degree of degeneracy, the set of equations read
\begin{subequations}\label{gs}
\begin{align}
&\frac{{\partial {n_s}}}{{\partial t}} + \frac{{\partial {n_s}{v_s}}}{{\partial x}} = 0,\\
&\frac{{\partial {v_s}}}{{\partial t}} + {v_s}\frac{{\partial {v_s}}}{{\partial x}} = \frac{{{e}}}{{{m_e}}}\frac{{\partial \phi }}{{\partial x}} - \frac{1}{{{m_e}}}\frac{{\partial {\mu}}}{{\partial x}} + \frac{{{\hbar ^2}}}{{2m_e^2}}\frac{\partial }{{\partial x}}\left( {\frac{1}{{\sqrt {{n_s}} }}\frac{{{\partial ^2}\sqrt {{n_s}} }}{{\partial {x^2}}}} \right),\\
&\frac{{{\partial ^2}\phi }}{{\partial {x^2}}} =  4\pi e \sum\limits_s {{n_s}},
\end{align}
\end{subequations}
in which the dependent variables, $n_s$ and $v_s$ refer to the number density and fluid velocity of given electron stream, indexed by $s$, and $\phi$ is the electrostatic potential. The last term in momentum equation arises due to quantum Bohm potential, by elimination of which the system (\ref{gs}) reduces to the classical Dawson's multistream model \cite{dawson}. Moreover, $\mu$ is the chemical potential of the electron gas which is related to the electron number density using an appropriate equation of state (EoS) and is used to close the hydrodynamic system (\ref{gs}). For isothermal electron gas of arbitrary degeneracy the EoS is
\begin{subequations}\label{eosr}
\begin{align}
&n(\nu ,T) = \frac{{{2^{7/2}}\pi {m_e^{3/2}}}}{{{h^3}}}{F_{1/2}}(\nu) =  - \frac{{{2^{5/2}}{{(\pi m_e{k_B}T)}^{3/2}}}}{{{h^3}}}{\rm{L}}{{\rm{i}}_{3/2}}[ - \exp (\nu )],\\
&P(\nu ,T) = \frac{{{2^{9/2}}\pi {m_e^{3/2}}}}{{3{h^3}}}{F_{3/2}}(\nu) =  - \frac{{{2^{5/2}}{{(\pi m_e{k_B}T)}^{3/2}}({k_B}T)}}{{{h^3}}}{\rm{L}}{{\rm{i}}_{5/2}}[ - \exp (\nu )],
\end{align}
\end{subequations}
in which $P$ is the statistical pressure of the gas and $\nu=\beta\mu$ with $\beta=1/k_B T$, where, $F_k$ is the Fermi integral of order $k$ given as
\begin{equation}\label{f}
{F_k}(\nu ) = \int_0^\infty  {\frac{{{x^k}}}{{\exp (x - \nu ) + 1}}} dx.
\end{equation}
In terms of polylog function, ${\rm{Li}}_{k}$, the Fermi integrals are defined as
\begin{equation}\label{pl}
{F_k}(\nu ) =  - \Gamma (k + 1){\rm{L}}{{\rm{i}}_{k + 1}}[ - \exp (\nu )],
\end{equation}
in which $\Gamma$ is the conventional gamma function. It is seen that, the thermodynamic identity, $\partial P/\partial\mu=n$, holds for the electron gas in the thermodynamic equilibrium. Note also that we ignore the chemical potential dependence on stream index in the gas and it is assumed that the index $s$ characterizes only the velocity spectrum in the system. For a multispecies plasmas, however, this index may apply to the chemical potential of species. The hydrodynamic model (\ref{gs}) may be cast into a more simple form as the effective Schr\"{o}dinger-Poisson model \cite{manfredi} for the system, using the Madelung transformations ${\cal N}_s(x,t) =\sqrt{n_s(x,t)}\exp[iS_s(x,t)/\hbar]$ and $v_s(x,t)=(1/m_e)\partial S_s(x,t)/\partial x$. By using the later definition, the continuity and momentum balance, after separation of real/imaginary parts become
\begin{subequations}\label{spn}
\begin{align}
&{m_e}\frac{{\partial {n_s}}}{{\partial t}} + \frac{{\partial {n_s}}}{{\partial x}}\frac{{\partial {S_s}}}{{\partial x}} + {n_s}\frac{{{\partial ^2}{S_s}}}{{\partial {x^2}}} = 0,\\
&\frac{{{\partial ^2}{S_s}}}{{\partial t\partial x}} + \frac{1}{m_s}\frac{{\partial {S_s}}}{{\partial x}}\frac{{{\partial ^2}{S_s}}}{{\partial {x^2}}} =  \frac{{{e}\partial \phi }}{{\partial x}} - \frac{{\partial {\mu}}}{{\partial x}} + \frac{{\partial {B_s}}}{{\partial x}},\\
&{B_s} = \frac{{{\hbar ^2}}}{{8{m_e}n_s^2}}\left[ {2{n_s}\frac{{{\partial ^2}{n_s}}}{{\partial {x^2}}} - {{\left( {\frac{{\partial {n_s}}}{{\partial x}}} \right)}^2}} \right],
\end{align}
\end{subequations}
which combining with ${\cal N}_s(x,t) =\sqrt{n_s(x,t)}\exp[iS_s(x,t)/\hbar]$ together with the Poisson's equation leads to the following Schr\"{o}dinger-Poisson system \cite{has}
\begin{subequations}\label{sp}
\begin{align}
&i\hbar{\frac{{\partial {\cal N}_s}}{{\partial t}} =  - \frac{{{\hbar ^2}}}{{2{m_e}}}\frac{{{\partial ^2}{\cal N}_s}}{{\partial {x^2}}} - e\phi {\cal N}_s + {\mu}{\cal N}_s},\\
&\frac{{{\partial ^2}\phi }}{{\partial {x^2}}} = 4\pi e\sum\limits_s {|{\cal N}_s{|^2}}.
\end{align}
\end{subequations}
For our purpose let us consider a particular case of multistream model in the linear perturbation limit in which every stream of electrons is a monoenergetic beam interacting with others through the electrostatic potential. One may linearize the system (\ref{nsp}) through the assumptions $n_s(x,t)=n_0+n_{1s}$, $\phi_s(x,t)=0+\phi_1$, $S_s(x)=S_0+p_s x$ and $\mu=\mu_0$ with $\mu_0$ being the equilibrium chemical potential of the gas and $p_s$ being the constant momentum of given stream. Also, $S_0$ results in a constant phase which is ignored in this analysis. In the steady state limit we are also able to decompose the state function into spatiotemporal product of variables as, ${\cal N}_s(x,t)=\psi_s(t)\psi_s(x)\exp(i k_s x)$, where, $\psi_s(x)=\sqrt{n_s(x)}$ with $k_s=p_s/\hbar$ being the de Broglie wavenumber of given stream. Hence, eliminating the first-order index, the normalized linear system of coupled equations read
\begin{subequations}\label{nsp}
\begin{align}
&\frac{{{d^2}{\Psi_s(x)}}}{{d{x^2}}} + 2i{k_s}\frac{{d{\Psi_s(x)}}}{{dx}} + \Phi(x)  + (E - k_s^2){\Psi_s(x)} = 0,\\
&\frac{{{d^2}\Phi(x) }}{{d{x^2}}} - \sum\limits_s f_s\Psi_s(x) = 0,
\end{align}
\end{subequations}
where $E=(\epsilon-\mu_0)/E_p$ is the normalized multistream system energy with $E_p=\hbar\omega_p$ being the plasmon energy and $\omega_p=\sqrt{4\pi e^2n_0/m_e}$ the electron plasma frequency. Also, $f_s$ represents the momentum ($\hbar k_s$) distribution function for given stream $s$ with the property $\sum\limits_s f_s=1$. However in current multistream model we have chosen $f_s=1$, for simplicity. Moreover, $\epsilon=\hbar\omega$ defines the common energy eigenvalue of the multistream system. Moreover, we used the normalization scheme, $\Psi_s(x)\to\Psi_s(x)/\sqrt{n_0}$, with $n_0$ being the equilibrium electron number density and $\Phi(x)=e\phi/E_p$. The space coordinate $x$ is normalized to the plasmon length, $l_p=2\pi/k_p$, with $k_p=\sqrt{2m_eE_p}/\hbar$ being the characteristic plasmon wavenumber. Therefore, the de Broglie wavenumber is normalized to the plasmon length and temperature to the plasmon temperature $T_p=E_p/k_B$. The system (\ref{nsp}), plus the temporal term proportional to $\exp(-i\omega t)$, describes the steady state evolution of an electron gas in the linear limit. To obtain the state functions ${\cal N}(x,t)=\sum\limits_s \Psi_s(x)\exp(-i\Omega t)$ ($\Omega=\omega/\omega_p$) and $\Phi(x)$ one has to evaluate $N$-coupled differential equations through the electrostatic potential each given for an electron stream. Note that $\Psi_s(x)$ characterize the pure states of the multistream system from which the mixed states are calculated. The fluid velocity of each stream satisfy the relation $v_s(x,t)=j_s(x,t)/n_s(x,t)$, where, $j_s(x,t)=i\hbar/(2m_e)[\partial{\cal N}_s(x,t)/\partial x\times {\cal N}^*_s(x,t)- \partial{\cal N}^*_s(x,t)/\partial x\times {\cal N}_s(x,t)]$ is the current density of given stream. This velocity is also given by the relation $v_s(x,t)=(\hbar/m_e){\rm{Im}}[\partial{\cal N}_s(x,t)/\partial x/{\cal N}_s(x,t)]$, which is identical with the pseudoparticle velocity in the pilot-wave theory for guiding equation. In this linear limit we have $v_s=\hbar k_s/m_e$. The multistream velocity is obtained through the state function as $v(x,t)=(\hbar/m_e){\rm{Im}}[\partial{\cal N}(x,t)/\partial x/{\cal N}(x,t)]$ by solving the $N$-coupled equations (\ref{nsp}).

\begin{figure}[ptb]\label{Figure1}
\includegraphics[scale=0.7]{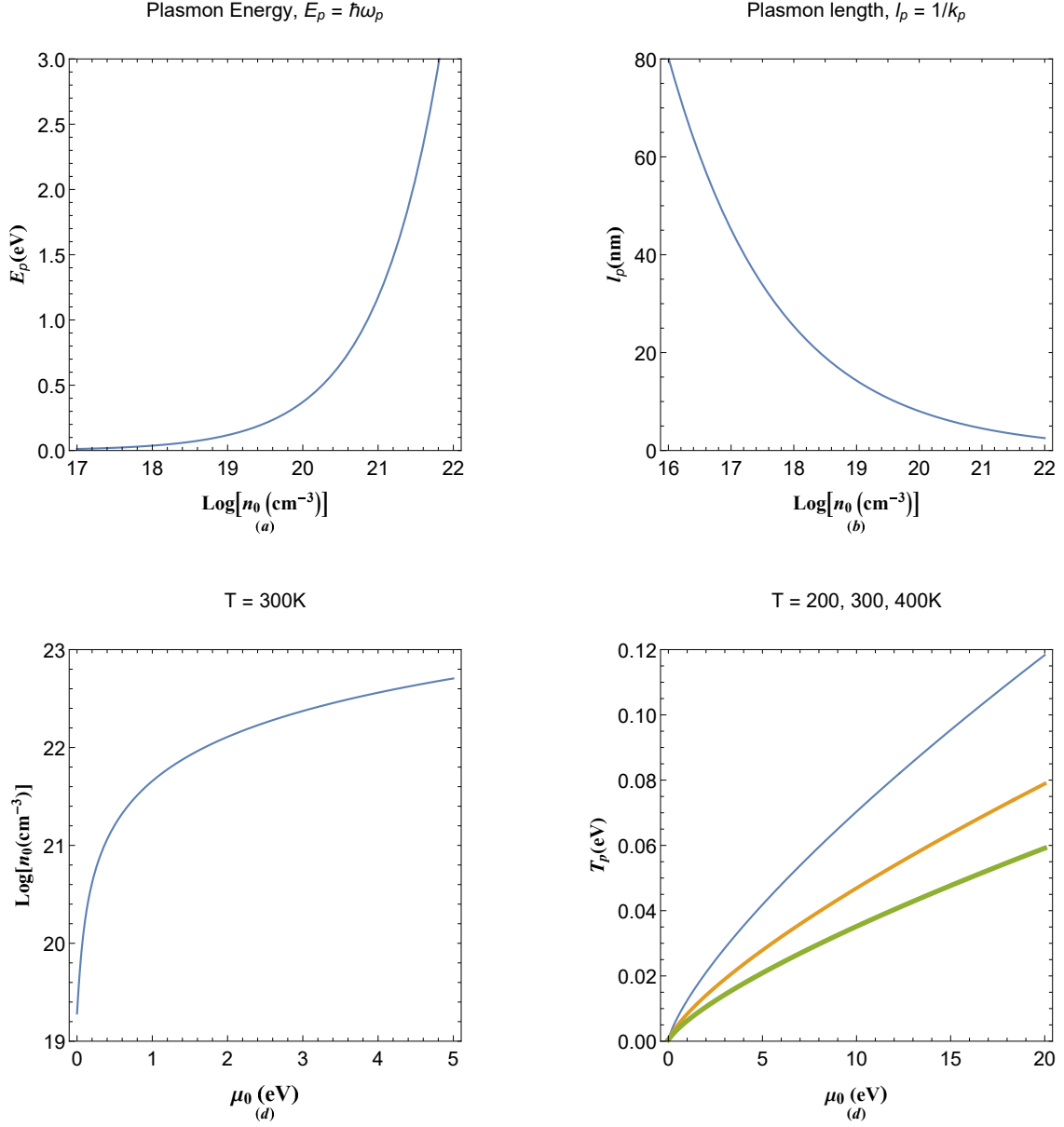}\caption{(a) Variation in the plasmon energy $E_p=\hbar\omega_p$ in terms of electron number density in logarithmic scale. (b) Variations in plasmon length $1/k_p$ with electron number density in logarithmic scale. (c) The electron concentration in terms of equilibrium chemical potential of arbitrary degenerate electron gas. (d) Variation of plasmon temperature $T_p=E_p/k_B$ in terms of the chemical potential of the electron gas for different values of the electron temperature. The increase in thickness of curves in plot (d) indicates the increase in varied parameter above the panel.}
\end{figure}

Figure 1 shows the variations in characteristic parameters of the plasmon system with electron number density and an arbitrary degenerate electron gas. In Fig. 1(a) it is shown that the plasmon energy varies up to few electronvolts from the classical to fully degenerate electron gas. Note that only in highly doped semiconductors and metals this energy becomes significant. The plasmon length variations in nanometer unit in terms of electron number density is shown in Fig. 1(b). This length sharply decreases with increase in number density to a fraction of a nanometer in a typical metal. The chemical potential variation is shown in Fig. 1(c). In the fully degenerate (zero temperature) limit in typical metals the chemical potential at $E=0$ or $\epsilon=\mu$ characterizes the fundamental Fermi energy level of the system which is assumed to be constant. Hence, current model is most appropriate for metals and nano-metallic density regime and beyond. Figure 1(d), on the other hand, shows the variation plasmon temperature, $T_p$ with the variations in chemical potential for different values of the electron temperature. It is remarked that, the plasmon temperature increases with increase in the chemical potential but decreases with increase in electron temperature.

\section{One-Stream Model and the Doppler Shift}

Despite the simplicity of the model (\ref{nsp}), it will be shown that it is useful in describing some fundamental physical phenomenon corresponding to the plasmon system. Consider a single stream ($s=1$) described by the following system
\begin{subequations}\label{s1}
\begin{align}
&\frac{{{d^2}{\Psi_1(x)}}}{{d{x^2}}} + 2i{k_1}\frac{{d{\Psi_1(x)}}}{{dx}} + \Phi(x)  + (E - k_1^2){\Psi_1(x)} = 0,\\
&\frac{{{d^2}\Phi(x) }}{{d{x^2}}} - \Psi_1(x) = 0,
\end{align}
\end{subequations}
satisfying the energy dispersion relation ${E} = {1}/{{{k^2}}} + {\left( {k + {k_1}} \right)^2}$ which reduces to the plasmon dispersion relation in the limit $k_1=0$. It is remarked that the particle branch of the energy dispersion is Doppler shifted due to the streaming electrons. However, the wave-like branch is not affected by the electron drift. It has been shown that in an inertial frame moving along with the electron beam the traveling wave solution to the system (\ref{s1}) becomes identical with that of the electron gas in rest frame \cite{akbtravel} with the beam speed replacing the energy eigenvalues.

\begin{figure}[ptb]\label{Figure2}
\includegraphics[scale=0.7]{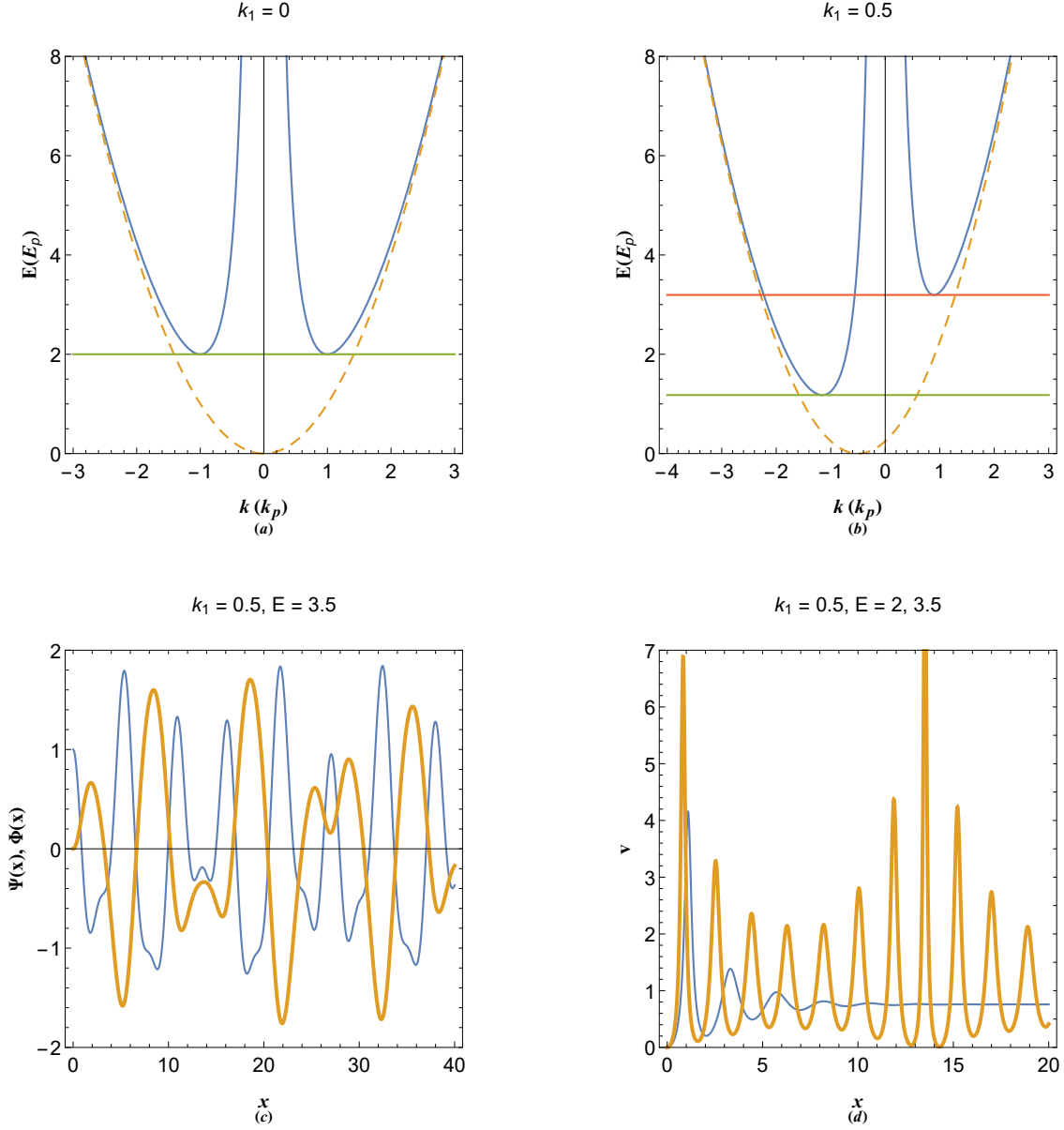}\caption{Dispersion curves of (a) static and (b) streaming free electron (dashed curve) and plasmon (solid curve) excitations. (c) Variation of state-functions $\Psi(x)$ (thin curve) and $\phi(x)$ (thick curve) in streaming electron gas with the de Broglie wavenumber $k_1=0.5$ at stable orbital $E=3.5$. (d) Quasiparticle velocity corresponding to the one-stream electron gas in (c) at stable $E=3.5$ (thick curve) and unstable $E=2$ (thin curve) orbital.}
\end{figure}

Figure 2 shows the energy dispersion and state functions of plasmon excitations for given parameters. The energy dispersion plasmon excitations for the case of $k_1=0$ (solid curves) along with the free electron dispersion (dashed curve) are shown in Fig. 2(a). There are stable plasmon excitations above the critical value $E=2$ (as shown by horizontal line) which are double-tone due to both particle-like ($k>1$) and wave-like ($k<1$) phenomena. However, below this critical line the excitation wavenumbers become complex and energy exchange occurs between the partcile-like and wave-like branches, as discussed in Ref. \cite{akbtravel}. Figure 2(b) depicts the energy dispersion of excitations for $k_1=0.5$. It is remarked that the free electron dispersion undergoes a Doppler shift and two critical minimum values for energy appear, namely, $E_{m1}\simeq 1.1786$ and $E_{m2}\simeq 3.1944$. For $E>E_{m2}$ the plasmon excitation with four real wavenumbers are stable. However, for $E_{m1}<E<E_{m2}$ only two of the excitations wavenumbers are real, hence, excitations are unstable. For $E<E_{m1}$ all four wavenumbers become complex and plasmon excitations become unstable again. However, there is a fundamental difference between the two unstable regimes $E_{m1}<E<E_{m2}$ and $E<E_{m1}$, as will be discussed later. Figure 2(c) shows the profiles of state functions, namely $\Psi(x)$ (thin curve) and $\Phi(x)$ (thick curve), for given stable oscillation parameter values. These state function have been obtained by numerical solution of (\ref{s1}) with initial conditions, $\Phi(0)=\Phi'(0)=\Psi'(0)=0$ and $\Psi(0)=1$. The variation of quasiparticle velocity for stable orbital ($E=3.5$ as thick curve) and unstable orbital ($E=2$ as thin curve) is depicted in Fig. 2(d). Evidently, there are oscillations in the velocity profiles which are damped for unstable energy orbital $E=2$.

\section{Two-Stream Model and Energy Band Formation}

Let us consider the following symmetric two stream system
\begin{subequations}\label{s2}
\begin{align}
&\frac{{{d^2}{\Psi _1}}}{{d{x^2}}} + 2i{k_1}\frac{{d{\Psi _1}}}{{dx}} + \Phi  + (E - k_1^2){\Psi _1} = 0\\
&\frac{{{d^2}{\Psi _2}}}{{d{x^2}}} + 2i{k_2}\frac{{d{\Psi _2}}}{{dx}} + \Phi  + (E - k_2^2){\Psi _2} = 0\\
&\frac{{{d^2}\Phi }}{{d{x^2}}} - \Psi _1 - \Psi _2 = 0,
\end{align}
\end{subequations}
where $k_1$ and $k_2$ are de Broglie wavenumbers of the streams. The energy dispersion relation can be obtained by assuming plane-wave expansions, $\Psi_1(x)=\Psi_{11}\exp(ikx)$, $\Psi_2(x)=\Psi_{21}\exp(ikx)$ and $\Phi_1(x)=\Phi_1\exp(ikx)$, which leads to the following eigenvalue system
\begin{equation}\label{ev2}
\left( {\begin{array}{*{20}{c}}
{E - {{(k + {k_1})}^2}}&0&1\\
0&{E - {{(k + {k_2})}^2}}&1\\
1&1&{{k^2}}
\end{array}} \right)\left( {\begin{array}{*{20}{c}}
{{\Psi _{11}}}\\
{{\Psi _{21}}}\\
{{\Phi _1}}
\end{array}} \right) = \left( {\begin{array}{*{20}{c}}
0\\
0\\
0
\end{array}} \right).
\end{equation}
The energy dispersion relation of two-stream system reads
\begin{equation}\label{dr}
E_\pm = {k^2} + \frac{1}{{{k^2}}} + \frac{1}{2}\left( {k_1^2 + k_2^2} \right) + k\left( {{k_1} + {k_2}} \right) \pm \frac{{\sqrt {4 + {k^4}{{\left( {{k_1} + {k_2}} \right)}^2}{{\left( {{k_1} + {k_2} + 2k} \right)}^2}} }}{{2{k^2}}}.
\end{equation}

\begin{figure}[ptb]\label{Figure3}
\includegraphics[scale=0.7]{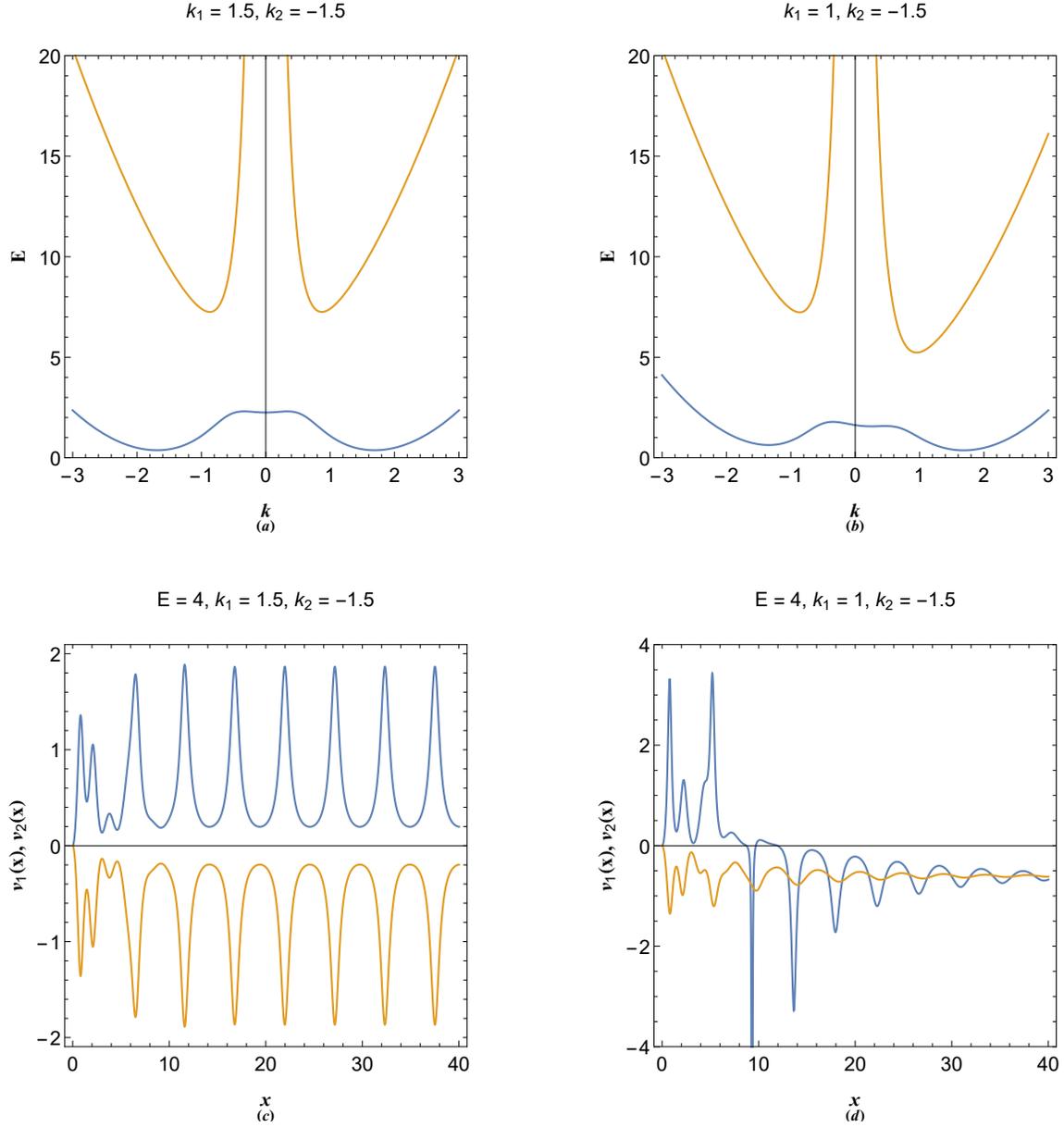}\caption{(a) Energy band structure of symmetric two-stream plasmon excitations. (b) Energy band structure of asymmetric two-stream plasmon excitations. Quasiparticle velocity of two-stream excitations in symmetric band gap. (d) Quasiparticle velocity of two-stream excitations in asymmetric band gap showing the phase mixing effect due to mode coupling of the energy bands.}
\end{figure}

It is remarked that, in the two-stream model extra energy band appears due to mode coupling between the streams. Figure 3 shows the structure of energy bands in two-stream model. In Fig. 3(a) the two streams have same but opposite velocity and the upper and lower energy bands are symmetric. The imbalanced two stream is shown in Fig. 3(b), the asymmetry of which is caused by the Doppler shift in particle-like branch. The quasiparticle orbital velocities in each stream are shown in Figs. 3(c) and 3(d) for symmetric and asymmetric cases. It is seen that in the asymmetric two stream model the orbital velocity of quasiparticle in electron beam with $k_1=1$ reverses at $E=4$ and merges with the other stream. The later phenomenon, which we may call the phase mixing effect, is a novel feature of the two-stream model caused by the collective wave-particle interactions in the energy band gaps. Note that in the preceding analysis (and the following) we consider equal density distribution for streams, for the sake of simplicity. In fact a generalized momentum distribution function, such as the Maxwell-Boltzmann, may be used in the Poisson's equation as weight function of stream probability functions $\Psi_N$. However, in the limit of full degeneracy the momentum distribution becomes unity.

\section{Energy Band Structure in Multistream Model}

To this end, it is straightforward to generalize the model to include a large number of streams each characterized by their de Broglie's wavenumber, $k_N$. Therefore, $N$-coupled differential equation system read
\begin{subequations}\label{sN1}
\begin{align}
&\frac{{{d^2}{\Psi _1}}}{{d{x^2}}} + 2i{k_1}\frac{{d{\Psi _1}}}{{dx}} + \Phi  + (E - k_1^2){\Psi _1} = 0,\\
&\frac{{{d^2}{\Psi _2}}}{{d{x^2}}} + 2i{k_2}\frac{{d{\Psi _2}}}{{dx}} + \Phi  + (E - k_2^2){\Psi _2} = 0,\\
&\hspace{10mm}\vdots\hspace{20mm}\vdots\hspace{20mm}\vdots\hspace{20mm}\\
&\frac{{{d^2}{\Psi _N}}}{{d{x^2}}} + 2i{k_N}\frac{{d{\Psi _N}}}{{dx}} + \Phi  + (E - k_N^2){\Psi _N} = 0,\\
&\frac{{{d^2}\Phi }}{{d{x^2}}} - \Psi _1 - \Psi _2 - \cdots - \Psi_N = 0,
\end{align}
\end{subequations}
Fourier analysis of (\ref{sN1}) leads to the following eigenvalue system
\begin{equation}\label{evn}
\left( {\begin{array}{*{20}{c}}
{E - {{(k + {k_1})}^2}}&0& \ldots &{\begin{array}{*{20}{c}}
{\begin{array}{*{20}{c}}
{}&{}&0&{\begin{array}{*{20}{c}}
{}&{}
\end{array}}
\end{array}}&1
\end{array}}\\
0&{E - {{(k + {k_2})}^2}}& \ldots &{\begin{array}{*{20}{c}}
{\begin{array}{*{20}{c}}
{}&{}&0&{\begin{array}{*{20}{c}}
{}&{}
\end{array}}
\end{array}}&1
\end{array}}\\
 \vdots & \vdots & \ddots &{\begin{array}{*{20}{c}}
{\begin{array}{*{20}{c}}
{}&{}&{\begin{array}{*{20}{c}}
 \vdots &{}
\end{array}}&{}
\end{array}}& \vdots
\end{array}}\\
{\begin{array}{*{20}{c}}
0\\
1
\end{array}}&{\begin{array}{*{20}{c}}
 \ldots \\
1
\end{array}}&{\begin{array}{*{20}{c}}
0\\
 \cdots
\end{array}}&{\begin{array}{*{20}{c}}
{\begin{array}{*{20}{c}}
{E - {{(k + {k_N})}^2}}&1
\end{array}}\\
{\begin{array}{*{20}{c}}
{}&{\begin{array}{*{20}{c}}
{}&1&{\begin{array}{*{20}{c}}
{}&{\begin{array}{*{20}{c}}
{}&{{k^2}}
\end{array}}
\end{array}}
\end{array}}
\end{array}}
\end{array}}
\end{array}} \right)\left( {\begin{array}{*{20}{c}}
{{\Psi _{11}}}\\
{\begin{array}{*{20}{c}}
{{\Psi _{21}}}\\
 \vdots \\
{{\Psi _{N1}}}
\end{array}}\\
{{\Phi _1}}
\end{array}} \right) = \left( {\begin{array}{*{20}{c}}
0\\
{\begin{array}{*{20}{c}}
0\\
 \vdots \\
0
\end{array}}\\
0
\end{array}} \right).
\end{equation}
The system (\ref{evn}) can be evaluated numerically to any number of streams in order to calculate the structure of energy bands in multistream electron system.

\begin{figure}[ptb]\label{Figure4}
\includegraphics[scale=0.7]{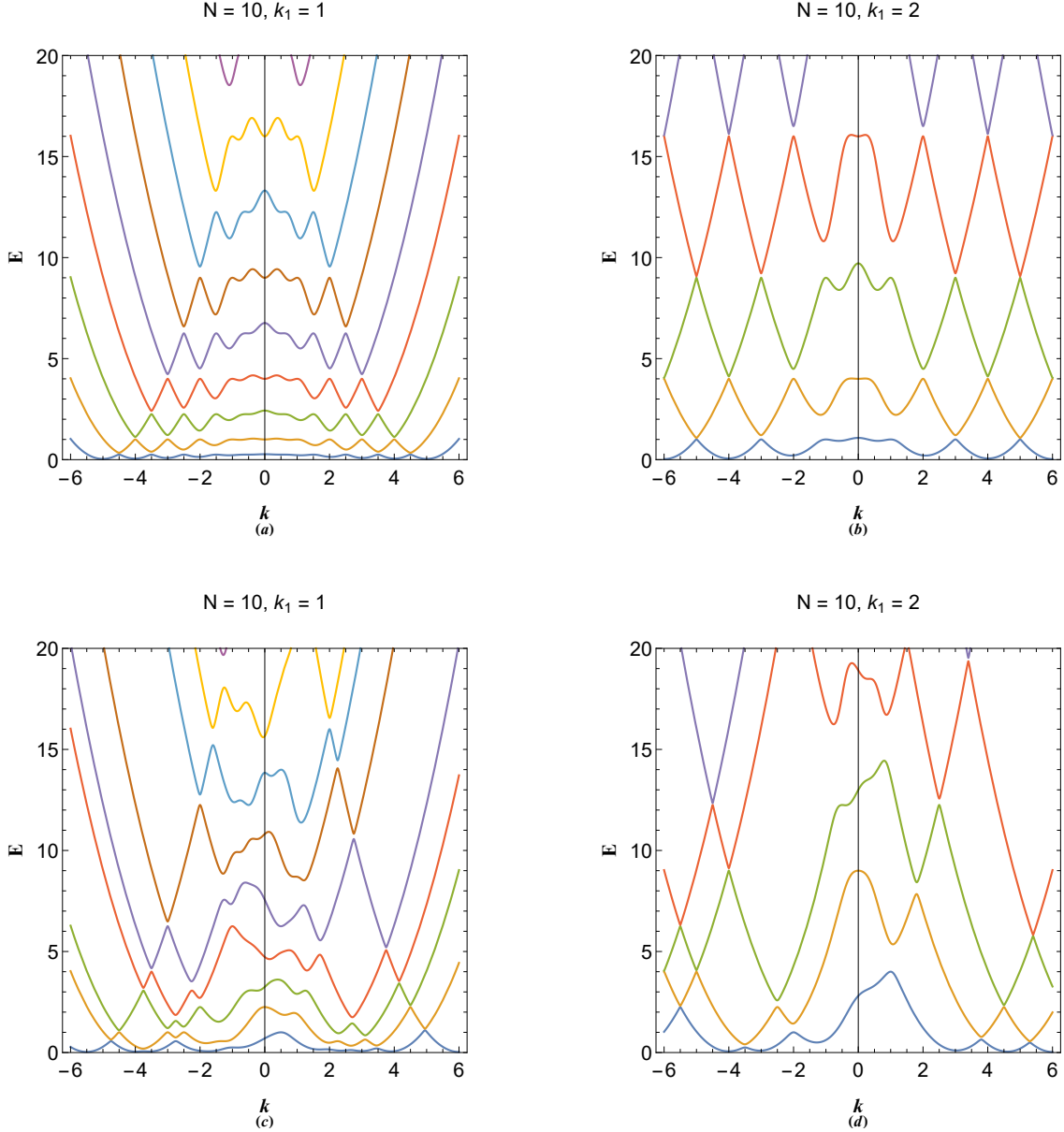}\caption{(a) The periodic $N$-coupled ($N=10$), $k_N=Nk_1$, energy band structure of multistream electron gas with $k_1=1$. (b) The periodic 10-coupled, $k_N=Nk_1$, energy band structure of multistream electron gas with $k_1=2$. (c) The random $N$-coupled ($N=10$), energy band structure of multistream electron gas with $k_1=1$. (d) The random 10-coupled energy band structure of multistream electron gas with $k_1=2$.}
\end{figure}

Figure 4 shows the electronic energy band structure of a 10-coupled ($N=10$) multistream system. The periodic stream profile in Fig. 4(a) corresponds to the especial case of $k_N=Nk_1$ with $N=10$ and $k_1=1$ illustrating different bands which are separated via band gaps. As mentioned previously the origin of band gaps in this case is the mode coupling between different electron streams, very similar to the electronic band structure in crystalline materials, as will be shown in the next section. Note that band gaps increase with increase in energy, $E$, but decrease with increase in wavenumber, $k$. The periodic 10-coupled band structure for $k_1=2$ is depicted in Fig. 4(b). It is remarked that increase in the stream speed leads to increase in the energy band widths. Figures 4(c) and 4(d) depict the band structure of 10-coupled randomly distributed multistream system in the arbitrary degenerate electron gas. The increase in the number of streams and arbitrary velocity distribution can lead to formation of a very complex energy band structure which describes the plasmon excitations in the multistream system.

We have already seen that presence of discrete electron streams leads to formation of energy band structure and energy gaps leading to complex wavenumbers with the imaginary part representing the growing/damping features. It has been shown \cite{akbdamp} that in collisionless electron systems the complex wavenumbers does not lead to dissipation of energy but the exchange between the particle-like and wave-like excitations very similar to the Landau damping phenomenon. In extreme limit of our multistream theory when every electron (velocity) specie constitute an individual stream with the weigh function, $f_s$, being the Maxwell-Boltzmann velocity distribution, large number of energy minibands form which are separated by tiny band gaps through which phase-mixing, i.e. quasiparticle scattering, can occur. In this case energy exchanges can take place between the collective electrostatic excitations and single electron oscillations by irreversible energy transfer from wave-like oscillations to the particle-like ones. the multistream model has been originally used by Dawson in order to give a physical interpretation of the Landau Damping effect \cite{dawson}. Indeed the effect can occur by resonant electrons which have speeds close to the plasmon excitation phase speed, i.e., $v\simeq E/\hbar k$, residing in a band gap.

To better understand the collisionless damping phenomenon, we consider the traveling wave solutions of a electron beam (stream) with normalized drift speed, $\gamma$, $\Phi(x-2\gamma t)$ and ${\cal N}(x,t)=\Psi(x-2\gamma t)\exp[i \gamma(x-2\gamma t)]$ obtained in Ref. \cite{akbtravel}, in which
\begin{equation}\label{wfb}
\left\{ {\begin{array}{*{20}{c}}
{{\Phi}(x-2\gamma t)}\\
{{\Psi}(x-2\gamma t)}
\end{array}} \right\} = \frac{1}{{2\alpha }}\left\{ {\begin{array}{*{20}{c}}
{{\Psi _0} + k_p^2{\Phi _0}}&{ - \left( {{\Psi _0} + k_w^2{\Phi _0}} \right)}\\
{ - \left( {{\Phi _0} + k_w^2{\Psi _0}} \right)}&{{\Phi _0} + k_p^2{\Psi _0}}
\end{array}} \right\}\left\{ {\begin{array}{*{20}{c}}
{\cos [{k_w}(x-2\gamma t)]}\\
{\cos [{k_p}(x-2\gamma t)]}
\end{array}} \right\},
\end{equation}
in which $\Phi_0$ and $\Psi_0$ define the initial values at $x=2\gamma t$ and the characteristic wave-like and particle-like wavenumbers $k_w$ and $k_p$ are given,respectively, as
\begin{equation}\label{eks}
{k_w} = \sqrt {(E_d - \alpha)/2},\hspace{3mm}{k_p} = \sqrt {(E_d + \alpha)/2},\hspace{3mm}\alpha  = \sqrt {{E_d^2} - 4},
\end{equation}
where $E_d=\gamma^2-\mu$ and $k_w k_p=1$. Inside the gap the wavenumbers become complex, i.e., $k=k_r+ik_i$ with $k_r$ and $k_i$ being the real and imaginary parts of the wavenumbers. It has been shown that \cite{akbtravel}, while the real parts are equal ($k_{wr}=k_{pr}$), the imaginary part of wave-like excitation is always negative, ($k_{wi}<0$) and that of the particle-like is always positive, ($k_{pi}=-k_{wi}>0$) for space-time range $x>2\gamma t$. Therefore, the wave-like/particle-like oscillations grow in space for energy values with imaginary de Broglie wavenumbers (inside the energy gaps) where the electron streams experience the so-called quantum drift instability \cite{akbtravel}. On the other hand, a close inspection of the solution (\ref{wfb}) reveals that the wave-like/particle-like oscillations experience damping in time, simultaneously. Generally speaking, particle-like/wave-like excitations of arbitrary degenerate electron beam always undergo spacial/temporal growing/damping inside the energy gaps. The above description of multistream electron behavior may be regarded as an elegant quantum description of the colissionless Landau damping effect due to the resonant wave-particle interactions and energy exchange between wave-like and particle-like oscillations within the energy band gaps. It is remarkable however that the stream velocity defined through, $v_s = {\mathop{\rm Im}\nolimits} [{\cal N}_x(x,t)/{{\cal N}}(x,t)]  = \gamma$, is invariant under the wave-particle processes, indicating total energy conservation. It can be shown that spacial/temporal growth/damping of wave-like/particle-like behavior is intrinsic behavior of an electron beam in quantum tunneling process where quantum drift instability takes place (similar to instability of electron stream excitations within the energy band gaps). The detailed study of the relationship between the wave-particle phenomenon and the collective tunneling through a potential barrier may be the subject of a future study in the framework of complex energy band structure and is beyond the scope of current research.

\section{Band Structure of 1D Plasmonic Crystals}

In this section we would like to generalize the theory of multistream model to band structure of plasmon excitations in periodic system like such as plasmonic crystals \cite{akbgap}. Considering a lattice of constant $a$ the crystal is characterized by reciprocal lattice vectors $G_N=NG_1$ with $G_1=2\pi/a$ being the first reciprocal lattice vector and $N$ is an integer number. Now, we model the electronic excitations through the $N$-coupled virtual streams (\ref{sN}) in which the reciprocal lattice wavevectors, $G_N$, play the role of de Broglie's wavenumber of virtual streams. Therefore, we have the following $N$-coupled virtual stream system with a solution of type ${\cal N}_N(x,t)=\Psi_N(x)\exp(iG_N x-i\Omega t+i\Theta_N)$ in which $G_N/2$ characterize the N-th Brillouin zone boundary and $\Theta_N$ is the arbitrary phase angle of the given stream.
\begin{subequations}\label{sN}
\begin{align}
&\frac{{{d^2}{\Psi _1}}}{{d{x^2}}} + 2iG_1\frac{{d{\Psi _1}}}{{dx}} + \Phi  + (E - G_1^2){\Psi _1} = 0,\\
&\frac{{{d^2}{\Psi _2}}}{{d{x^2}}} + 2iG_2\frac{{d{\Psi _2}}}{{dx}} + \Phi  + (E - G_2^2){\Psi _2} = 0,\\
&\hspace{10mm}\vdots\hspace{20mm}\vdots\hspace{20mm}\vdots\hspace{20mm}\\
&\frac{{{d^2}{\Psi _N}}}{{d{x^2}}} + 2i{G_N}\frac{{d{\Psi _N}}}{{dx}} + \Phi  + (E - G_N^2){\Psi _N} = 0,\\
&\frac{{{d^2}\Phi }}{{d{x^2}}} - \Psi _1 - \Psi _2 - \cdots - \Psi_N = 0,
\end{align}
\end{subequations}
the Fourier analysis of (\ref{sN}) leads to the following eigenvalue system
\begin{equation}\label{evnc}
\left( {\begin{array}{*{20}{c}}
{E - {{(k + {G_1})}^2}}&0& \ldots &{\begin{array}{*{20}{c}}
{\begin{array}{*{20}{c}}
{}&{}&0&{\begin{array}{*{20}{c}}
{}&{}
\end{array}}
\end{array}}&1
\end{array}}\\
0&{E - {{(k + {G_2})}^2}}& \ldots &{\begin{array}{*{20}{c}}
{\begin{array}{*{20}{c}}
{}&{}&0&{\begin{array}{*{20}{c}}
{}&{}
\end{array}}
\end{array}}&1
\end{array}}\\
 \vdots & \vdots & \ddots &{\begin{array}{*{20}{c}}
{\begin{array}{*{20}{c}}
{}&{}&{\begin{array}{*{20}{c}}
 \vdots &{}
\end{array}}&{}
\end{array}}& \vdots
\end{array}}\\
{\begin{array}{*{20}{c}}
0\\
1
\end{array}}&{\begin{array}{*{20}{c}}
 \ldots \\
1
\end{array}}&{\begin{array}{*{20}{c}}
0\\
 \cdots
\end{array}}&{\begin{array}{*{20}{c}}
{\begin{array}{*{20}{c}}
{E - {{(k + {G_N})}^2}}&1
\end{array}}\\
{\begin{array}{*{20}{c}}
{}&{\begin{array}{*{20}{c}}
{}&1&{\begin{array}{*{20}{c}}
{}&{\begin{array}{*{20}{c}}
{}&{{k^2}}
\end{array}}
\end{array}}
\end{array}}
\end{array}}
\end{array}}
\end{array}} \right)\left( {\begin{array}{*{20}{c}}
{{\Psi _{11}}}\\
{\begin{array}{*{20}{c}}
{{\Psi _{21}}}\\
 \vdots \\
{{\Psi _{N1}}}
\end{array}}\\
{{\Phi _1}}
\end{array}} \right) = \left( {\begin{array}{*{20}{c}}
0\\
{\begin{array}{*{20}{c}}
0\\
 \vdots \\
0
\end{array}}\\
0
\end{array}} \right).
\end{equation}
Note that we assumed an empty lattice approximation in which the atomic lattice potential is negligible compared to the plasmon energy. Such assumption can be valid in fully degenerate regime due to effective charge screening or in the case on weak-potential plasmonic lattices. The system (\ref{evnc}) may be evaluated numerically for a finite number of Brillouin zone approximation. Analytical solution to energy band dispersion exists for $3$-coupled system with lattice momentum $-G$, $0$ and $+G$. For $N=3$ approximation-order, one obtains
\begin{subequations}\label{dr3}
\begin{align}
&{E_1} = \frac{1}{{3{k^2}}}\left[ {3 + 2{G^2}{k^2} + 3{k^4} - \frac{\tau}{\delta } - \delta } \right],\\
&{E_2} = \frac{1}{{6{k^2}}}\left[ {2\left( {3 + 2{G^2}{k^2} + 3{k^4}} \right) + \frac{\tau }{\delta }\left( {1 + {\rm{i}}\sqrt 3 } \right) + \delta \left( {1 - {\rm{i}}\sqrt 3 } \right)} \right],\\
&{E_3} = \frac{1}{{6{k^2}}}\left[ {2\left( {3 + 2{G^2}{k^2} + 3{k^4}} \right) + \frac{\tau }{\delta }\left( {1 - {\rm{i}}\sqrt 3 } \right) + \delta \left( {1 + {\rm{i}}\sqrt 3 } \right)} \right],\\
&\delta  = {\left[ {{G^6}{k^6} - 36{G^4}{k^8} + \frac{1}{2}\sqrt {{{\left( {54 - 2{G^6}{k^6} + 72{G^4}{k^8}} \right)}^2} - 4{{\tau}^3}} } - 27 \right]^{1/3}},\\
&\tau = {9 + {G^4}{k^4} + 12{G^2}{k^6}}.
\end{align}
\end{subequations}

\begin{figure}[ptb]\label{Figure5}
\includegraphics[scale=0.7]{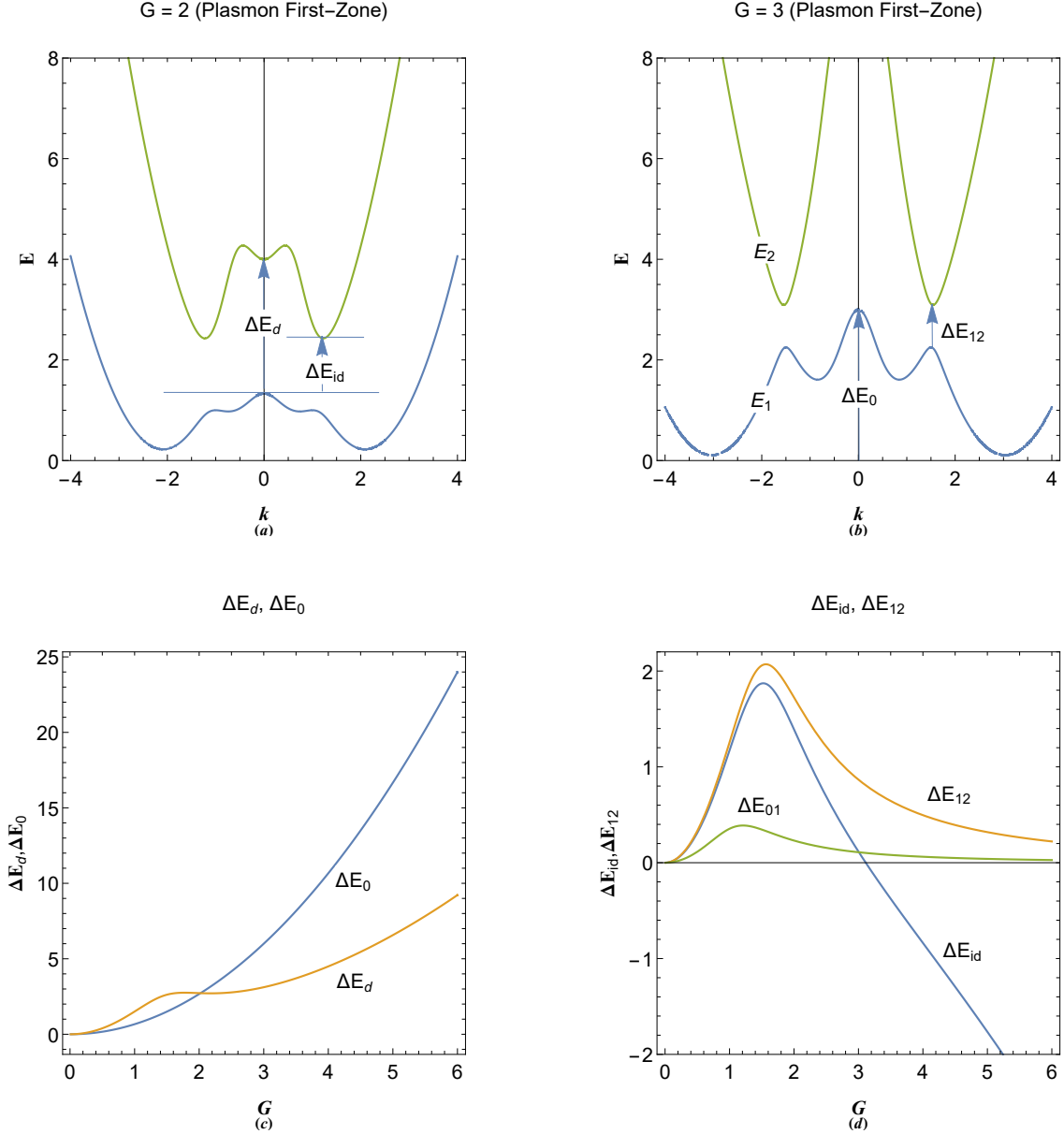}\caption{(a) The electronic band structure in 1D plasmonic lattice with $G=2$ in three-zone ($N=3$) empty-lattice approximation showing direct and indirect band gaps. (b) The electronic band structure in 1D plasmonic lattice with $G=3$ in three-zone ($N=3$) empty-lattice approximation showing the band gap at first Brillouin zone boundary, $k=G/2$. (c) Variations of $\Delta E_0$ and $\Delta E_{d}$ in terms of the reciprocal lattice vector, $G=2\pi/a$, where, $a$ is the lattice constant. (d) Variations of $\Delta E_{id}$, $\Delta E_{12}$ and the first conduction band height $\Delta E_{01}$ in terms of the reciprocal lattice vector, $G=2\pi/a$.}
\end{figure}

Figure 5 depicts the results of calculation for $N=3$ plasmonic crystal lattice dispersion in empty lattice approximation. It is remarkable that the band gap is still present in the absence of lattice potential due to the collective electron interactions contrary to the free electron model of solids \cite{kit}. Figure 5(a) shows the energy band structure with $G=2$. Direct and indirect band gaps are evident which are the result of mode coupling between different virtual streams. There are in fact three band the upper one not shown in the figure. Figure 5(b) shows the band structure for $G=3$ and the band gap $\Delta E_{12}$ between the energy bands $E_1$ and $E_2$ at the first Brillouin zone, $k=G/2$. In terms of reciprocal lattice vector $G$ one obtains $\Delta E_d=3G^2/2$ for the direct gap at $k=0$, and
\begin{subequations}\label{eid}
\begin{align}
&\Delta E_{id} = \frac{1}{{3{G^2}}}\left[ {12 + \frac{{7{G^4}}}{4} + \frac{{\left( {1 - {\rm{i}}\sqrt 3 } \right)\left( {36 + {G^8}} \right)}}{\zeta } + \left( {1 + {\rm{i}}\sqrt 3 } \right)\zeta } \right],\\
&\Delta E_{12} = \frac{{108 - 36{\rm{i}}\sqrt 3  + \left( {3 - {\rm{i}}\sqrt 3 } \right){G^8} + 3{\zeta ^2} + {\rm{i}}\sqrt 3 {\zeta ^2}}}{{3{G^2}\zeta }},\\
&\zeta  = {\left( {6{\rm{i}}{G^4}\sqrt 3 \sqrt {36 - 4{G^4} + {G^8}} - {G^{12}}}  - 216 \right)^{1/3}},
\end{align}
\end{subequations}
for the corresponding indirect gap and direct gap at first Brillouin zone, in three zone, $N=3$, approximation. Interesting features appear for variations of these gap with the reciprocal lattice vector (lattice constant). Figure 5(c) shows that $\Delta E_0$ increases monotonically with increase in $G$ and consequently decrease in lattice constant $a$. However, with increase of $G$ the value of direct gap at $k=0$ first increases and reaches a maximum value at $G\simeq 1.7726$ and then passes through a minimum value at $G\simeq 2.17572$. Figure 5(d) depicts variations in the indirect gap and the direct gap size at first Brillouin zone boundary. It is remarked that the indirect gap maximizes at $G\simeq 1.52319$ and closes at $G\simeq 3.11302$. The value of $\Delta E_{12}$ maximizes at $G\simeq 1.56508$. The first plasmon conduction band $\Delta E_{01}$ which occurs at $k=G$ is an important gap for fully degenerate electron gas (at zero temperature), where all the electrons are packed under the Fermi level ($E=0$ or $\epsilon=\mu_0$). The variation of this gap in terms of the reciprocal lattice vector is shown also in Fig. 5(d). It is noticed that the gap maximized at the value $G\simeq 1.20944$. Analytical expression for this gap for $N=3$ is
\begin{subequations}\label{e01}
\begin{align}
&\Delta {E_{01}} = \frac{1}{{6{G^2}}}\left[ {6 + 10{G^4} - \frac{{2\left( {9 + 13{G^8}} \right)}}{\delta } - 2\delta } \right],\\
&\delta  = {\left[ {\frac{1}{2}\sqrt {{{\left( {54 + 70{G^{12}}} \right)}^2} - 4{{\left( {9 + 13{G^8}} \right)}^3}} - 35{G^{12}}} - 27\right]^{1/3}}.
\end{align}
\end{subequations}

\begin{figure}[ptb]\label{Figure6}
\includegraphics[scale=0.7]{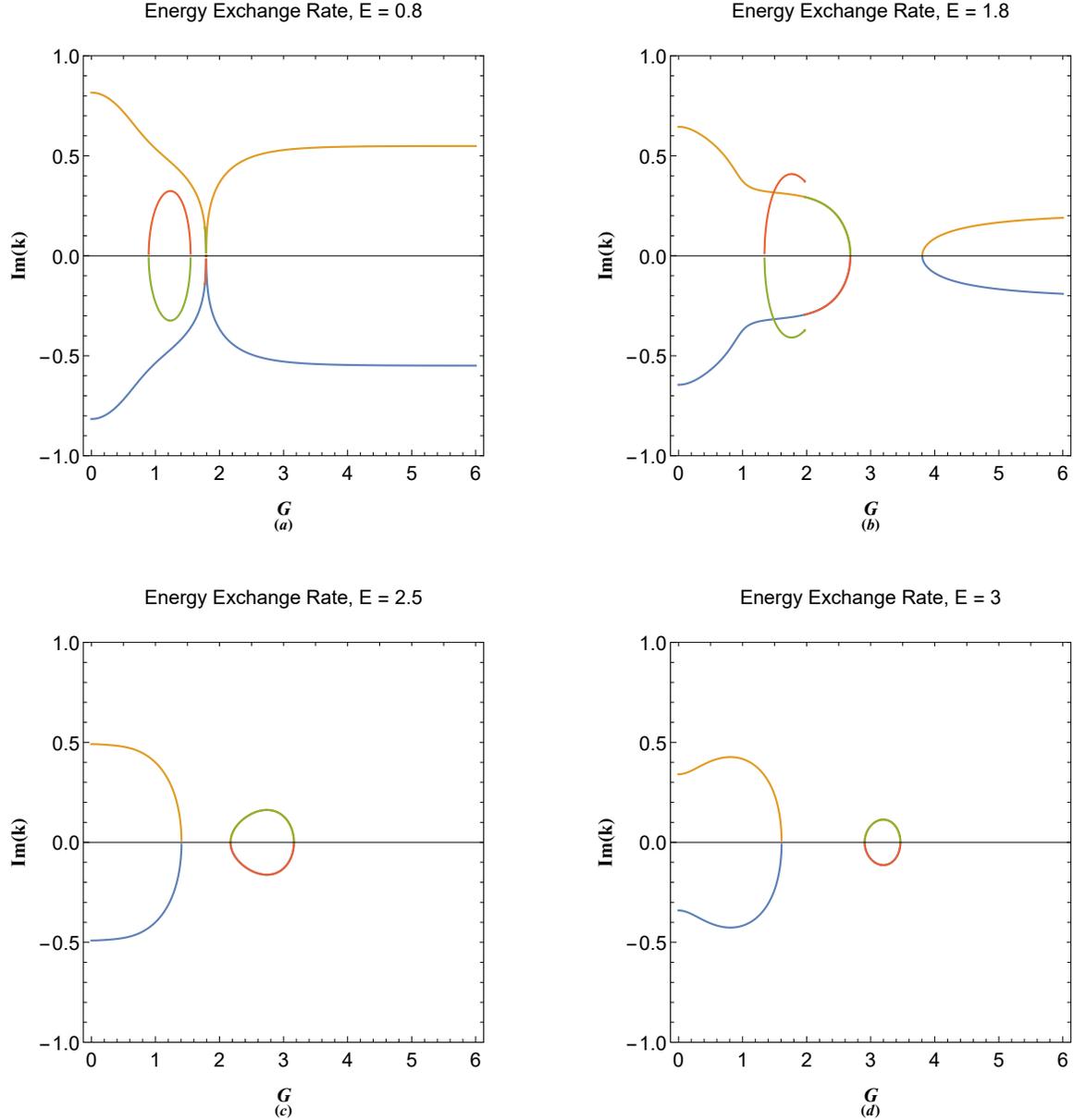}\caption{Variations in imaginary parts of plasmon excitation wavenumbers in plasmonic crystal in terms of reciprocal lattice vector $G$ at orbital (a) $E=0.8$ (b) $E=1.8$ (c) $E=2.5$ and (d) $E=3$. The possive branches are small wavelength particle-like and negative large wavelength wave-like plasmon excitation wavenumbers corresponding to each energy orbital.}
\end{figure}

Figure 6 shows the variations in imaginary wavenumber in terms of reciprocal lattice vector for different energies. The symmetric nature of figure indicates that excitations are dissipation free but energy exchange occurs between particle-like and wave-like oscillations. The imaginary part of wavevector at the band gaps play important role in Zener tunneling phenomenon in semiconductor diodes \cite{ash}. Recently, the propagation of a single stream electron beam studied in Ref. \cite{akbtravel}, reveals that the plasmonic excitations in the intrinsic energy gap in plasmonic excitations, due to wave-particle branch coupling below the critical value $E<2E_p$, leads to spacial growing/damping of the wave-like/particle-like excitations. It is however, concluded that the collective wave-particle interactions of dual-nature plasmonic excitations inside energy band gaps is accompanied by enhancement of wave-like amplitude and reduction in particle-like one through the space. This is a novel aspect of collective quantum interaction phenomenon detailed investigation of which is required in a future research. The positive/negative branches of imaginary components in Fig. 6 are particle-like/wave-like damp/growth rates. These imaginary wavenumbers always appear due to coupling of a wave-like excitation dispersion branch with that of a particle-like which leads to the appearance of energy gaps between separate bands.

\begin{figure}[ptb]\label{Figure7}
\includegraphics[scale=0.7]{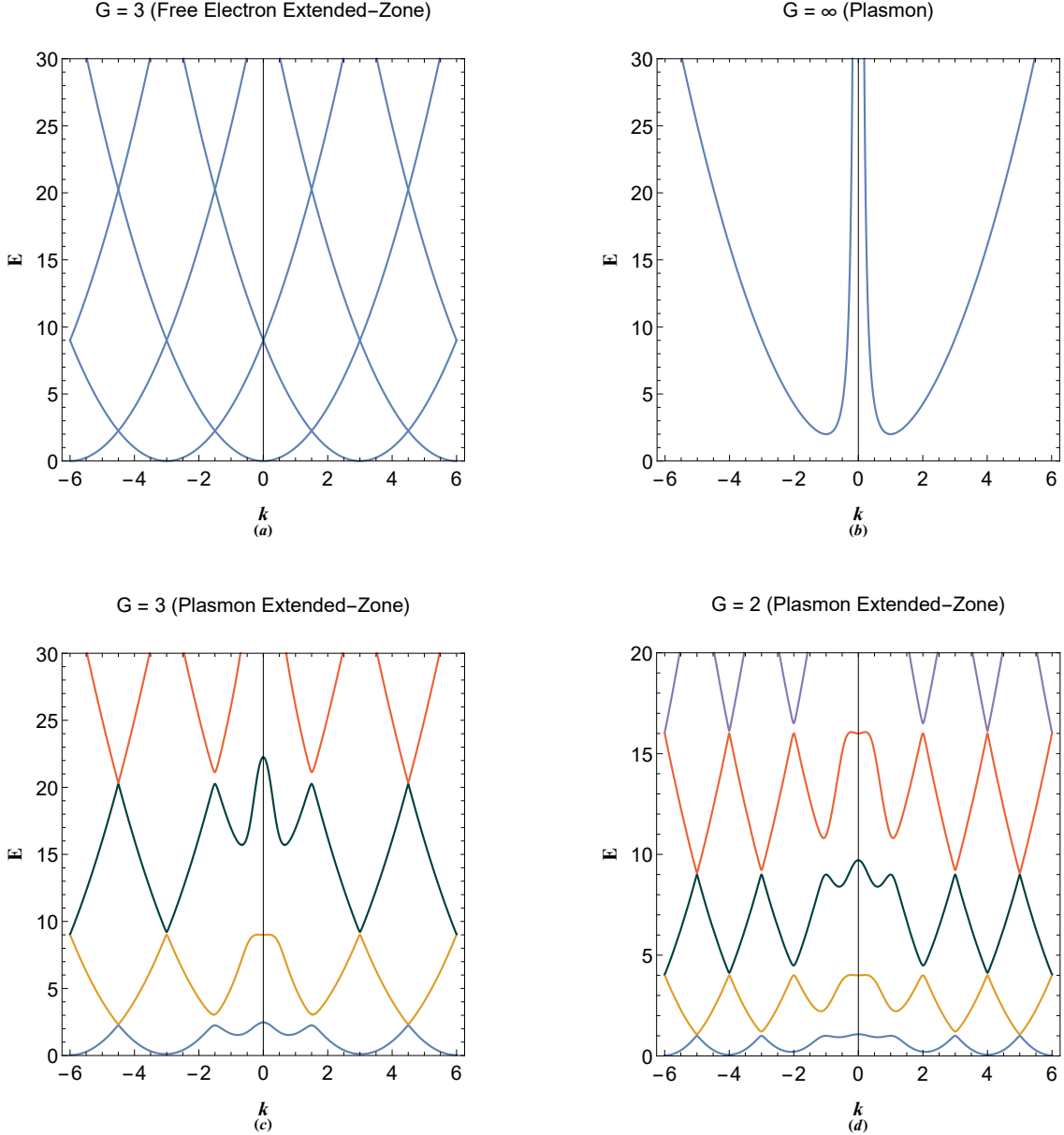}\caption{(a) The electronic extended-zone energy band structure in free electron model. (b) The plasmon excitation band in the limit $G\to\infty$. (c) The extended-zone electronic band structure of plasmonic crystal with $G=3$. (d) The extended-zone electronic band structure of plasmonic crystal with $G=2$.}
\end{figure}

Figure 7 shows the calculation results for band structure in 10-coupled approximation ($N=10$). Our result in Fig. 7(c) may be directly compared to the empty lattice band structure of free electron model with $G=3$ shown in Fig. 7(a). It is remarked that the band gap in our model occurs through collective electronic excitations even in the absence of lattice potential. The free electron lattice band structure may be obtained in our model by setting the coupling electrostatic field $\phi$ to zero. Figure 7(b) shows limiting case of $G\gg 1$ which is obviously the energy dispersion curve for ordinary plasmon excitation in electron gas. Figures 7(c) and 7(d) show the band structure for different values of $G$. It is remarked that band gaps take place at the Brillouin zone boundary the size of which decreases with increase in $k$ but increase with increase in $E$. The exact similarity between Figs. 7(c) and 4(b) shows that the problem of excitations in plasmonic crystals and quantum multistream have the same root. As a further generalization one may consider the effect of dynamic ions \cite{akbion} on the plasmonic lattice or super-lattice structure.

\begin{figure}[ptb]\label{Figure8}
\includegraphics[scale=0.7]{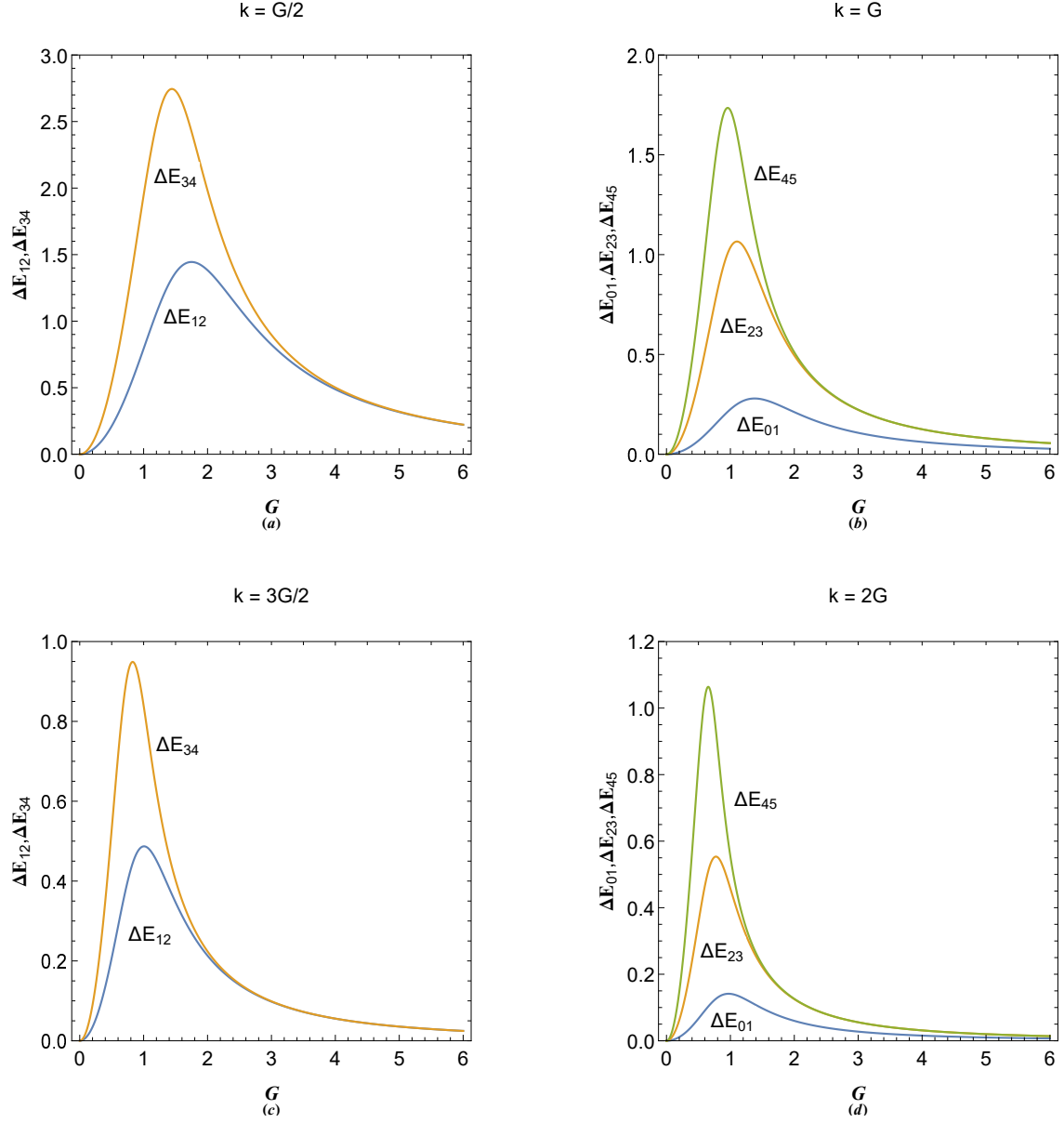}\caption{(a) Variation of the first two gap size at first Brilloun-zone boundary ($k=G/2$) with the reciprocal lattice vector size, $G=2\pi/a$ with $a$ being the lattice constant. (b) Variation of the first conduction band and first two higher gap sizes at $k=G$ with the reciprocal lattice vector size, $G=2\pi/a$. (c) Variation of the first two gap size at second Brilloun-zone boundary ($k=3G/2$) with the reciprocal lattice vector size, $G=2\pi/a$. (d) Variation of the first conduction band and two higher gap size at $k=2G$ with the reciprocal lattice vector size, $G=2\pi/a$.}
\end{figure}

In Fig. 8 we have shown the variations in band gaps in various Brillouin zone boundaries with respect to the reciprocal lattice vector. Figure 8(a) shows the gap size variation at first zone boundary at $k=G/2$ for $\Delta E_{12}=E_2-E_1$ and $\Delta E_{34}=E_4-E_3$. It is seen that $\Delta E_{12}$ and $\Delta E_{34}$, respectively, maximize at $G\simeq 1.74903$ and $G\simeq 1.44203$, that is, the gaps maximize at values of $G$ which tend to decrease with increase of the energy bands. Comparing Figs. 5(d) and 8(a), it is also noted that, the size of the band gap $\Delta E_{12}$ strongly depends on the number of lattice sites in $N$-coupled approximation in the plasmonic crystal. The band gap variations at $k=G$ are shown in Fig. 8(b). The first conduction band, $\Delta E_{01}=E_1$ ($E=0$ is the Fermi level for fully degenerate electron gas at zero temperature), at $k=G$ maximizes at the value $G=1.37936$. It is seen that the gap sizes are larger for upper bands, while, relatively lower in size with respect to those of the first zone. It is noted that $\Delta E_{23}$ and $\Delta E_{45}$ at $k=3G/2$, respectively, maximize at $G\simeq 1.1031$ and $G\simeq 0.9581$. Figures 8(c) and 8(d) reveal that gaps at higher zone boundaries maximize at lower $G$-values. It is noted that $\Delta E_{12}$ and $\Delta E_{34}$, respectively, maximize at $G\simeq 1.0061$ and $G\simeq 0.8282$. Also, $\Delta E_{01}$, $\Delta E_{12}$ and $\Delta E_{34}$ at $k=2G$, respectively, maximize at $G\simeq 0.9709$, $G\simeq 0.7738$ and $G\simeq 0.6537$. The variation of band structure in terms of lattice parameter may play a fundamental role in effective plasmonic crystal for technology. Note that the reciprocal lattice vector is related to the equilibrium electron number-density through $n_0=1/a^3=G^3/8\pi^3$ where $G$ is normalized with respect to the plasmon wavevector $k_p=\sqrt{2m_e\omega_p/\hbar}$ with $\omega_p=\sqrt{4\pi e^2 n_0/m_e}$ being the characteristic plasmon frequency. For instance, Fig. 8(b) shows that the first conduction band gap maximizes for $G\simeq 1.4$ (in $k_p$ unit), i.e., for a equilibrium electron density of $n_0=0.00368568n_p\simeq 1.28\times 10^{22}$cm$^{-3}$ with $n_p=\pi^3e^6 m_e^3/16\hbar^6$ being the plasmon density, which is slightly larger than the number density of a typical metallic elements \cite{kit} and is very close to the critical screening point $E_p=2E_F$ \cite{akbground}. On the other hand, the maximum value of the valence-conduction gap is $\Delta E_{01}\simeq 0.3$ (in $E_p$ unit). As an example the plasmon energy of metallic Sodium is $E_p\simeq 5.9$eV giving the energy gap size of $\Delta E_{01}\simeq 1.77$eV at $k=G$. The corresponding gap size at $k=2G$ is $\Delta E_{01}\simeq 0.7$eV and becomes much smaller at higher electron momentum for large $N$ at boundaries $k=NG$. Note that one has to take into account the dynamic effects of lattice ions \cite{akbion} which leads to sinking the conduction band into the Fermi electron sea. The calculated amount of ion potential effect on the energy band gap in the first-order perturbation approximation is known to be constant \cite{kit} and independent of the number $N$ at boundaries $k=NG$.

\section{Electron-Lattice Binding Effect}

In this section we would like to study the effect of electronic binding to lattice sites on the energy band structure of plasmon excitations, in the empty lattice approximation. To this end, we consider the following normalized and linearized non-Hermitian system, which includes the spacial damping effect. After the separation of spatiotemporal variables, one obtains
\begin{subequations}\label{sps}
\begin{align}
&\frac{{{\partial ^2}{{\cal M}_G}}}{{\partial {x^2}}} + 2\xi \frac{{\partial {{\cal M}_G}}}{{\partial x}} + \Phi {{\cal M}_G} + E {{\cal M}_G} = 0,\\
&\frac{{{\partial ^2}\Phi }}{{\partial {x^2}}} + 2\xi \frac{{\partial {\Phi}}}{{\partial x}} - \sum\limits_G {|{\cal M}_G|} = 0,
\end{align}
\end{subequations}
where, $\xi$ denotes the strength of plasmon oscillation damping, due to the electronic binding to the periodic lattice sites. However, in this simplified model, we do not want to go into details of the dependence of the damping parameter on other electron gas parameters, such as the equilibrium electron number-density and temperature. It is evident that, the system (\ref{sps}) should admit the general solution, ${\cal M}_G(x)=\Psi_N(x)\exp(iNGx-\xi|x-Na|+i\Theta_N)$, with the time dependent solution as ${\cal N}_G(x,t)={\cal M}_G(x)\exp(-i\Omega t)$, where the functions $\Psi_G(x)$ and $\Phi(x)$ satisfy the following $N$-coupled system
\begin{subequations}\label{sNs}
\begin{align}
&\frac{{{d^2}{\Psi _1}}}{{d{x^2}}} + 2iG_1\frac{{d{\Psi _1}}}{{dx}} + \Phi  + (E - G_1^2 - \xi^2){\Psi _1} = 0,\\
&\frac{{{d^2}{\Psi _2}}}{{d{x^2}}} + 2iG_2\frac{{d{\Psi _2}}}{{dx}} + \Phi  + (E - G_2^2 - \xi^2){\Psi _2} = 0,\\
&\hspace{10mm}\vdots\hspace{20mm}\vdots\hspace{20mm}\vdots\hspace{20mm}\\
&\frac{{{d^2}{\Psi _N}}}{{d{x^2}}} + 2i{G_N}\frac{{d{\Psi _N}}}{{dx}} + \Phi  + (E - G_N^2 - \xi^2){\Psi _N} = 0,\\
&\frac{{{d^2}\Phi }}{{d{x^2}}} - \Psi _1 - \Psi _2 - \cdots - \Psi_N = 0,
\end{align}
\end{subequations}
Fourier analysis of which results in the following eigenvalue system
\begin{equation}\label{evncs}
\left( {\begin{array}{*{20}{c}}
{E - {Q_1}}&0& \ldots &{\begin{array}{*{20}{c}}
{\begin{array}{*{20}{c}}
{}&{}&0&{\begin{array}{*{20}{c}}
{}&{}
\end{array}}
\end{array}}&1
\end{array}}\\
0&{E - {Q_2}}& \ldots &{\begin{array}{*{20}{c}}
{\begin{array}{*{20}{c}}
{}&{}&0&{\begin{array}{*{20}{c}}
{}&{}
\end{array}}
\end{array}}&1
\end{array}}\\
 \vdots & \vdots & \ddots &{\begin{array}{*{20}{c}}
{\begin{array}{*{20}{c}}
{}&{}&{\begin{array}{*{20}{c}}
 \vdots &{}
\end{array}}&{}
\end{array}}& \vdots
\end{array}}\\
{\begin{array}{*{20}{c}}
0\\
1
\end{array}}&{\begin{array}{*{20}{c}}
 \ldots \\
1
\end{array}}&{\begin{array}{*{20}{c}}
0\\
 \cdots
\end{array}}&{\begin{array}{*{20}{c}}
{\begin{array}{*{20}{c}}
{E - {Q_N}}&1
\end{array}}\\
{\begin{array}{*{20}{c}}
{}&{\begin{array}{*{20}{c}}
{}&1&{\begin{array}{*{20}{c}}
{}&{\begin{array}{*{20}{c}}
{}&{{k^2}}
\end{array}}
\end{array}}
\end{array}}
\end{array}}
\end{array}}
\end{array}} \right)\left( {\begin{array}{*{20}{c}}
{{\Psi _{11}}}\\
{\begin{array}{*{20}{c}}
{{\Psi _{21}}}\\
 \vdots \\
{{\Psi _{N1}}}
\end{array}}\\
{{\Phi _1}}
\end{array}} \right) = \left( {\begin{array}{*{20}{c}}
0\\
{\begin{array}{*{20}{c}}
0\\
 \vdots \\
0
\end{array}}\\
0
\end{array}} \right),
\end{equation}
where $Q_N=(k+G_N)^2-\xi^2$.

\begin{figure}[ptb]\label{Figure9}
\includegraphics[scale=0.7]{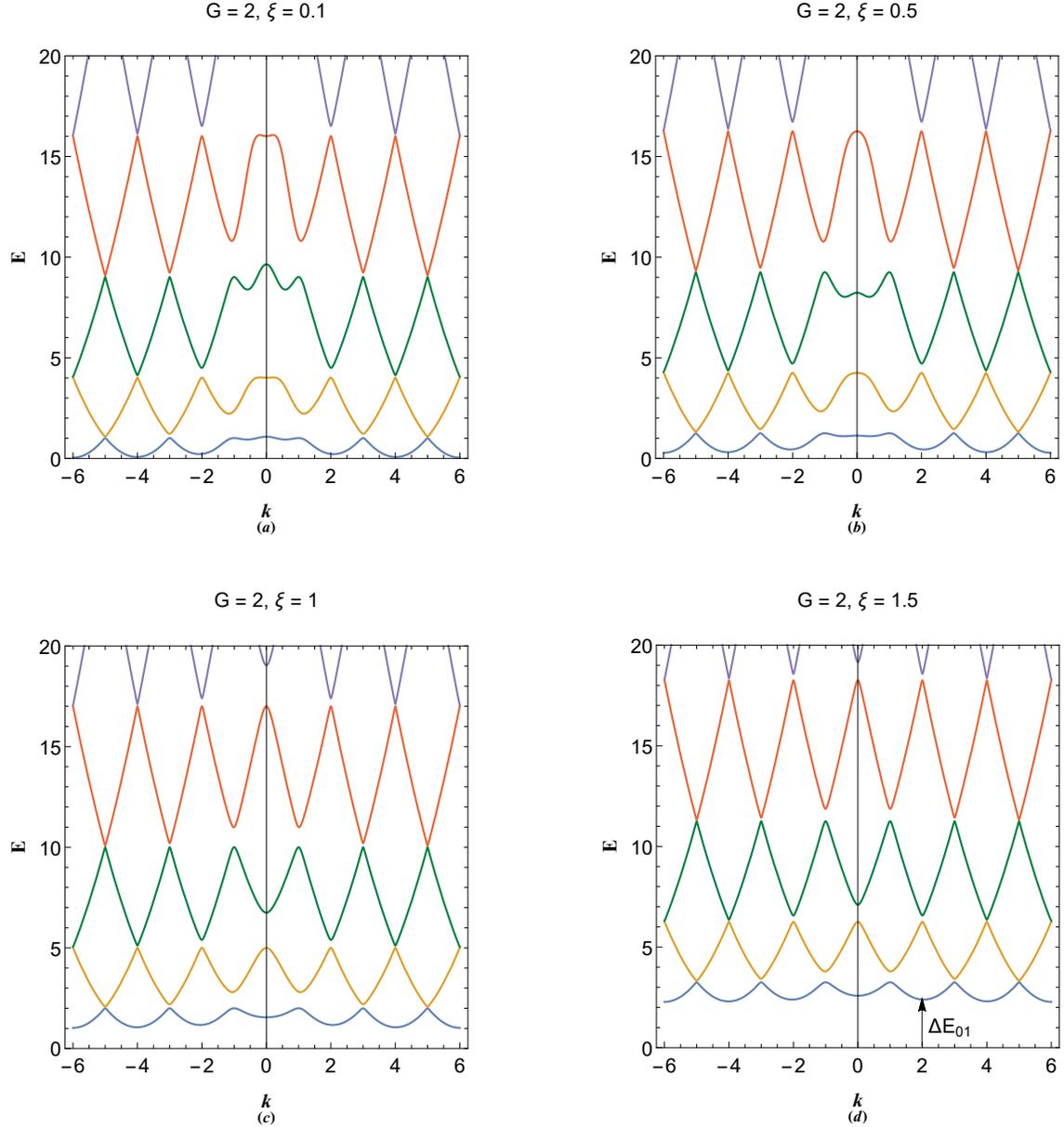}\caption{Variation in the electronic band structure of plasmon excitations in a plasmonic crystal with $G=2\pi/a$ for (a) $\xi=0.1$, (b) $\xi=0.5$, (c) $\xi=1$ and (d) $\xi=1.5$.}
\end{figure}

Figure 9 shows the effect of electron-lattice binding on the plasmonic band structure of periodic system with $G=2$ and different values of the binding strength parameter, $\xi$. It is clearly remarked that the increase in binding strength leads to the overall shift of plasmonic excitation bands to higher energies. It is further remarked that the energy gaps at the long wavelength limit $k=0$ decrease sharply by increase of the binding parameter. This is due to the significant effect of the electron-lattice binding on the wave-like branch rather than the particle-like one. It is also clearly remarked that the ground state electronic valence-conduction gap, $\Delta E_{01}$ at $k=NG$, through which the valence electrons can tunnel in the nearly free electron model, becomes smaller for larger values of the electron momentum, $\hbar k$. For higher binding strength regime ($\xi>1$), shown in Fig. 9(d), which we call the tight-binding limit where the electrons are tightly bound to the lattice sites, the first conduction plasmon energy band shift to much higher energies with the band inaccessible to Fermi electrons at $E=0$ ($\epsilon=\mu_0$) at zero temperature limit, thus, leading to insulating solid-state plasmon gap. Therefore, the critical value of binding parameter, $\xi$, may provide a quantitative measure for the Mott metal-insulator transition phenomenon in terms of the ground state gap energy, $\Delta E_{01}$, at zero temperature limit, where electro-hole process can occur. At finite temperature, on the other hand, electrons can excite to much higher energy bands and collective phenomenon become more pronounced.

\begin{figure}[ptb]\label{Figure10}
\includegraphics[scale=0.7]{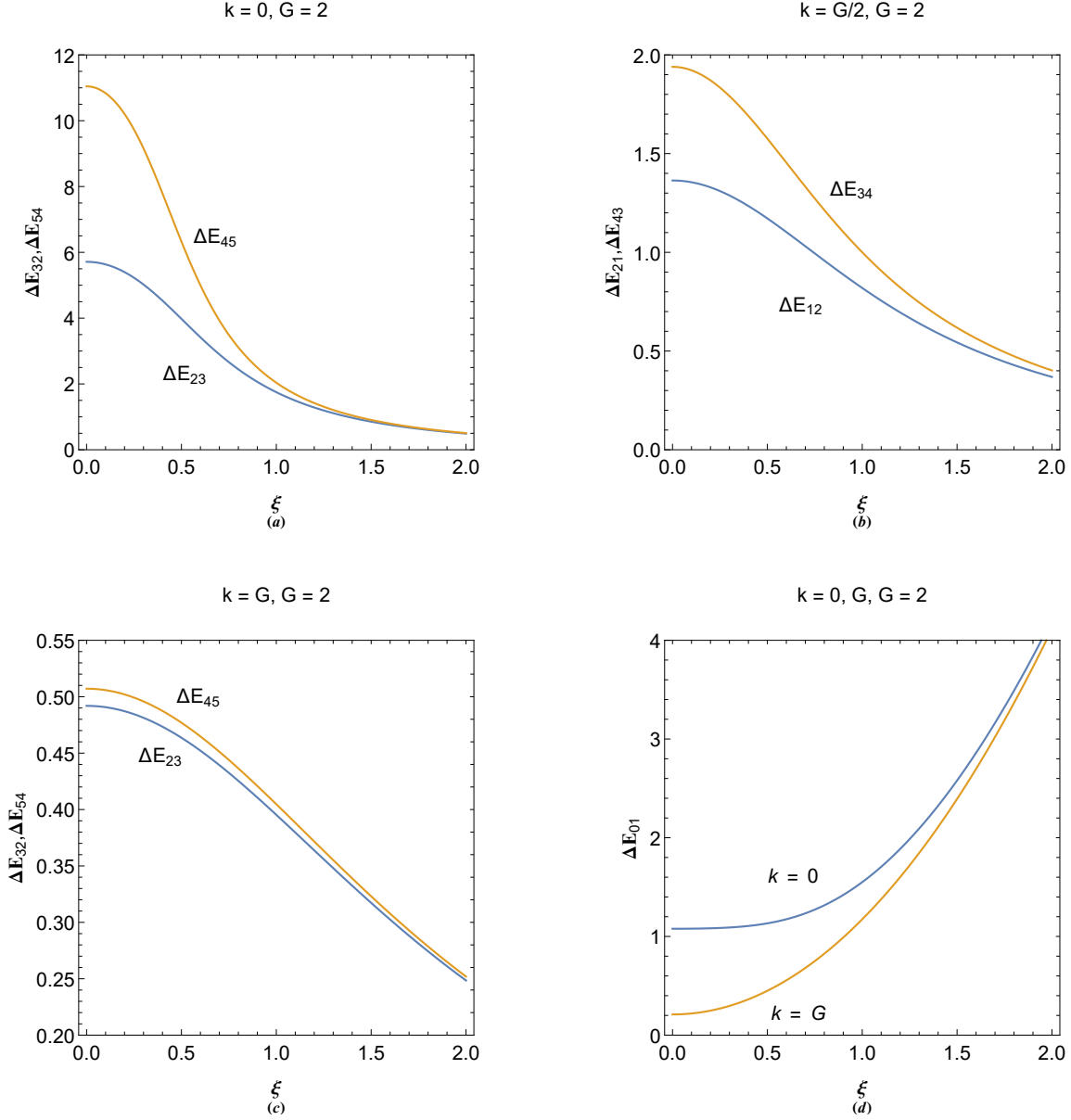}\caption{(a) Variation of the first two gap size at $k=0$ with the reciprocal lattice vector size, $G=2$. (b) Variation of the first two gap size at first Brilloun-zone boundary ($k=G/2$) with the reciprocal lattice vector size, $G=2$. (c) Variation of the first two gap size at secong Brilloun-zone boundary ($k=G$) with the reciprocal lattice vector size, $G=2$. (d) Variation of the conduction band height, $\Delta E_{10}$, for $k=0,G$ with the reciprocal lattice vector size, $G=2$.}
\end{figure}

Figure 10 shows the variations in various band gap sizes for 1D plasmonic crystal with reciprocal lattice vector size $G=2$ in terms of the electron binding parameter. Figure 10(a) shows the long wavelength ($k=0$) gap size variation. It is remarked that, with increase in the binding parameter the energy gaps $\Delta E_{45}$ and $\Delta E_{23}$ decrease sharply and saturate to the same value for large $\xi$, which is also clearly evident from Fig. 9. Figure 10(b) shows that the energy gaps $\Delta E_{34}$ and $\Delta E_{12}$ also decrease with increase in the value of $\xi$, but, with lower rate compared to gaps at $k=0$. The variations in the gaps $k=G$ of Fig. 10(c) becomes very small, indicating that the energy gaps at larger $k$ are less affected by the binding parameter variations. Variation in the first plasmon conduction bands $\Delta E_{01}$ for values of wavevectors $k=0,G$ are shown in Fig. 10(d). It is seen that $\Delta E_{01}$ at $k=0$ and $k=G$ increase with increase in $\xi$ and become identical for large $\xi$. It is concluded that in the tight-binding limit, $\xi\gg 1$, the band gaps tend to close and we obtain a free electron-like dispersion, similar to Fig. 7(a), with a very large ground state gap $\Delta E_{01}\gg 1$.

\section{Plasmon-Phonon Coupling Effect}

As a final remark, we would like to consider the effect of heavy species like dynamic ions on the band structure of 1D plasmonic crystals. In such a case we have a $N+1$-coupled system which may be written as
\begin{subequations}\label{sNi}
\begin{align}
&\frac{{{d^2}{\Psi _1}}}{{d{x^2}}} + 2iG_1\frac{{d{\Psi _1}}}{{dx}} + \Phi  + (E - G_1^2){\Psi _1} = 0,\\
&\frac{{{d^2}{\Psi _2}}}{{d{x^2}}} + 2iG_2\frac{{d{\Psi _2}}}{{dx}} + \Phi  + (E - G_2^2){\Psi _2} = 0,\\
&\hspace{10mm}\vdots\hspace{20mm}\vdots\hspace{20mm}\vdots\hspace{20mm}\\
&\frac{{{d^2}{\Psi _N}}}{{d{x^2}}} + 2i{G_N}\frac{{d{\Psi _N}}}{{dx}} + \Phi  + (E - G_N^2){\Psi _N} = 0,\\
&\sigma\frac{{{d^2}{\Psi _i}}}{{d{x^2}}} - \Phi  + (E + \mu){\Psi _i} = 0,\\
&\frac{{{d^2}\Phi }}{{d{x^2}}} - \Psi _1 - \Psi _2 - \cdots - \Psi_N + \Psi_i = 0,
\end{align}
\end{subequations}
where $\Psi_i$ denotes the ion wavefunction and $\sigma=m_e/m_i$ is the electron to ion mass ratio. As before, the Fourier analysis of the $N+1$-coupled system (\ref{sNi}) leads to the desired energy bands.

\begin{figure}[ptb]\label{Figure11}
\includegraphics[scale=0.7]{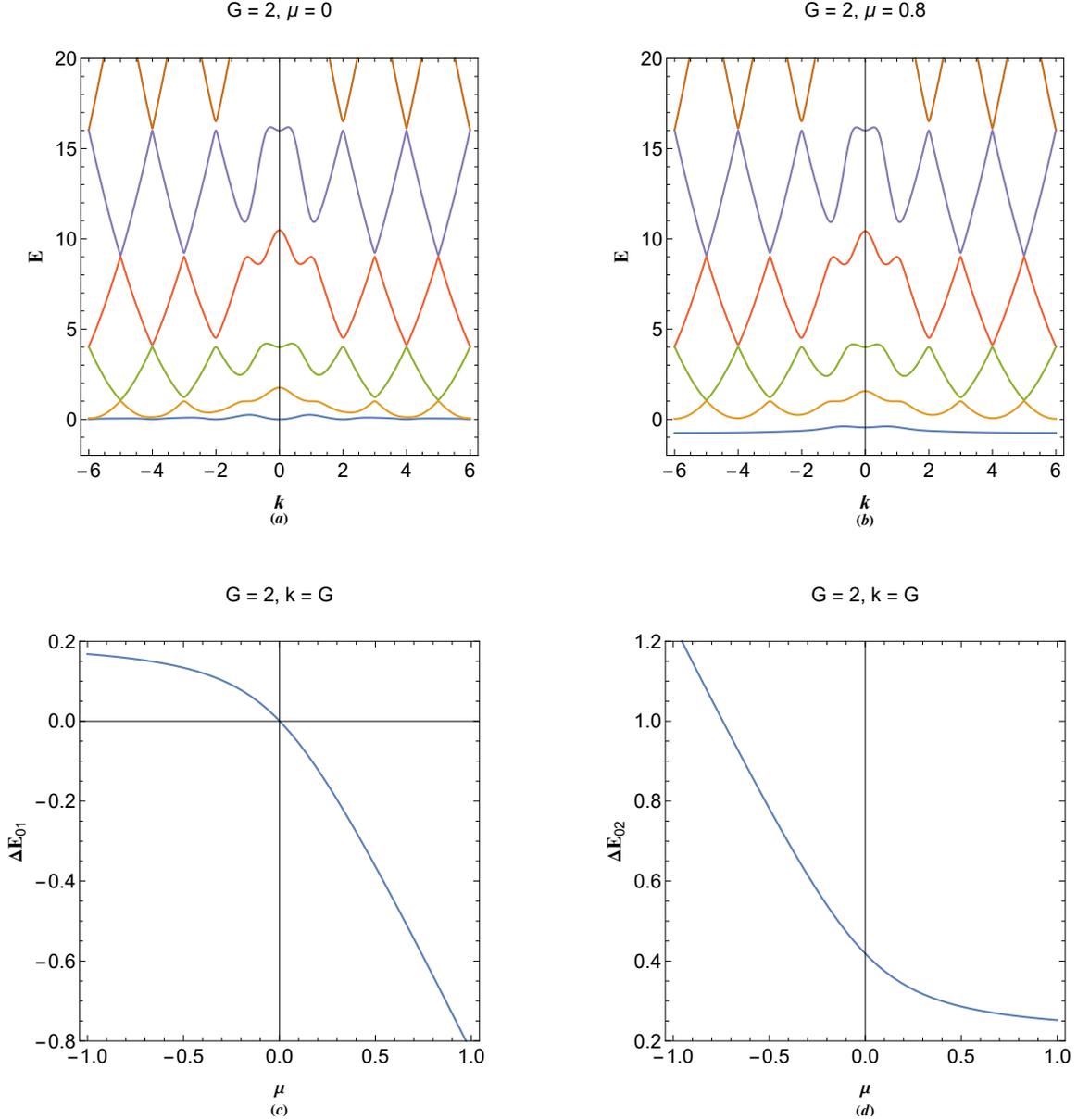}\caption{Band structure of 1D plasmonic crystal with $G=2$ in the presence of dynamic ions for (a) classical electron gas with $\mu=0$ and (b) degenerate electron gas with $\mu=0.8$. The variation in the gap between (c) first band and (d) second band from top of Fermi level with the change in normalized chemical potential.}
\end{figure}

Figure 11 shows the band structure for given values of the chemical potential and reciprocal lattice vector. The effect of heavy species such as dynamic ions on the energy dispersion of plasmonic excitations has recently been studied in \cite{akbion}. For the electron gas in semiconductor regime with $\mu=0$, as shown in Fig. 11(a), it is remarked that a nearly flat band (phonon-like low energy band) appears at $E=0$ (the Fermi level). The flatness of a conduction band is an indication of decreased mobility of electrons which is caused by electrostatic coupling of electrons to inertial ions. However, existence of such band is important for long wavelength phonon-assisted plasmon excitations in metals in the zero temperature limit. For a fully degenerate electron gas with $\mu=0.8$, the lowest energy band resides well below the Fermi energy level embracing a large amount of nearly free electrons in the gas, as seen in Fig. 11(b). The variation in gap between first band and the top of Fermi level $\Delta E_{01}$ is shown in Fig. 11(c) in terms of the normalized chemical potential. It is seen that the first band at $k=G$ touches the Fermi level at exact value of $\mu=0$ above/under which value the first band is belove/above the Fermi energy level. Moreover, Fig. 11(d) depicts the gap of second band from Fermi level at $k=G$ for $G=2$. The second band never touches the Fermi level for the chemical potential values in semiconductors and metallic density regime. However, the second energy band gap from $E=0$ decreases as the chemical potential increases. It should be noted that plasmonic crystals, unlike ordinary solids, can constitute from different charged species other. Also, the lattice sites can be interfaces between different plasmonic (metallic and semiconductor) layers, so called superlattice configuration, which do not contribute electrostatic potential to energy band structure. Therefore, current empty lattice model may well apply to a wide range of plasmonic crystal and superlattice configurations.

We have already considered the multistream electrostatic systems which only include electrons and ions. However, the simplified current model may be further generalized to include multispecies complex plasmas with a wide range of mass and charge-state spectrum or even gravitationally coupled uncharged quantum fluids. As discussed earlier, the energy band gap structure formation in quantum multistream system is expected be the origin of collisionless quantum stream instability and Landau damping effects. Therefore it is concluded that, these effects not only are characteristics of electrostatic systems, but also are inevitable in uncharged mass/spin multistream systems coupled through gravitational/magnetic potentials. The fundamental difference between electrostatic and gravitational Landau damping is that for gravitational case the damping occurs for wavenumbers larger than the Jeans wavenumber below which Jeans instability occurs.

\section{Conclusion}

We used the multifluid model to study the plasmonic excitations in electron gas with arbitrary degree of degeneracy by reducing the quantum hydrodynamic model into the $N$-coupled pseudoforce system. The energy band structure of a multistream system was obtained by linearizing the coupled differential equations which indicated that the energy bands form due to discrete stream filaments in the system and mode coupling by collective electrostatic interactions. Such velocity filaments may also be the root to collisionless damping and stream instability by gap opening very similar to the crystalline solids. Current model, generalized to virtual streams, was used to calculate the electronic band structure in one-dimensional plasmonic crystal. The dependence of energy band gaps on the lattice spacing is also studied in detail. The electronic band structure of a electron system can have essential effect on many characteristics of collective excitations in plasmonic crystals and metallic superlattices. We further studied the effect electron-lattie binding effect on the energy band structure of plasmonic crystals which indicates that with increase in the strength of the electron binding the first energy conduction band shifts to higher values where unaccessible to electrons at the Fermi sea. Inclusion of dynamic inertial ions in the plasmonic crystals, on the other hand, reveals that for degenerate electrons a flat-like ground state energy band appears inside the Fermi sea of free electrons due to electrostatic interaction between free electrons and ions decreasing the electron mobility substantially at this level. Therefore, current model of plasmonic excitations is capable of incorporating a wide range realistic features of electron dynamics in one-dimensional periodic structures and show akin similarities in band structure between multistream electron gas and plasmonic crystals.

\section{Data Availability}

The data that support the findings of this study are available from the corresponding author upon reasonable request.

\end{document}